\documentclass{aastex}
\usepackage{emulateapj5}

\slugcomment{To appear in the AJ (2001 Oct issue)}

\begin{document}

\title{MSX, 2MASS, and the LMC: A Combined Near and Mid Infrared View}
\author{Michael P. Egan}
\affil{Air Force Research Laboratory, Space Vehicles Directorate,
Hanscom AFB, MA  01731} 
\author{Schuyler D. Van Dyk}
\affil{Infrared Processing and Analysis
Center/Caltech, 100-22, Pasadena, CA  91125} 
\author{Stephan D. Price}
\affil{Air Force Research Laboratory, Space Vehicles Directorate,
Hanscom AFB, MA  01731}

\begin{abstract}
The Large Magellanic Cloud (LMC) has been observed by both the Midcourse
Space Experiment (MSX) in the mid-infrared and the Two Micron All Sky Survey
(2MASS) in the near-infrared. We have performed a cross-correlation of the
1806 MSX catalog sources and nearly 1.4 million 2MASS catalogued point and
extended sources and find 1664 matches. Using the available color
information, we identify a number of stellar populations and nebulae,
including main sequence stars, giant stars, red supergiants, carbon- and
oxygen-rich asymptotic giant branch (AGB) stars, planetary nebulae, H II
regions, and other dusty objects likely associated with early-type stars.
731 of these sources have no previous identification. We compile a listing
of all objects, which includes photometry and astrometry. The 8.3 $\mu $m
MSX sensitivity is the limiting factor for object detection: only the
brighter red objects, specifically the red supergiants, AGB stars, planetary
nebulae and HII regions, are detected in the LMC. The remaining objects are
likely in the Galactic foreground. The spatial distribution of the infrared
LMC sources may contribute to understanding stellar formation and evolution
and the overall galactic evolution. We demonstrate that a combined mid- and
near-infrared photometric baseline provides a powerful means of identifying
new objects in the LMC for future ground-based and space-based follow-up
observations.
\end{abstract}

\section{Introduction}

Among our nearest galactic neighbors, the Magellanic Clouds, with their
relative proximity, low line-of-sight extinction, lower mean metallicity,
and large numbers of stellar populations, can provide us with vital
information about our own Milky Way and galactic evolution in general. A
multitude of studies at a large variety of wavelengths have been conducted
on the Clouds for decades. Large-scale optical surveys, such as the
Magellanic Clouds Survey (Zaritsky, Harris, \& Thompson 1997), are compiling
results on a large number of stellar objects and nebulae. To obtain a view
particularly of the post-main-sequence stellar populations in these galaxies
one needs to turn to the infrared (IR). Two recent infrared surveys have
provided a wealth of new data on the Large Magellanic Cloud (LMC) in the
wavelength regime from 1.2 to 25 $\mu$m. At near-IR wavelengths, the Two
Micron All Sky Survey (2MASS) has detected nearly 2 million stars in the $10%
\arcdeg\times10\arcdeg$ area also surveyed by the Midcourse Space Experiment
(MSX) mid-IR survey.

MSX provides an improvement over previous mapping of the LMC in the mid-IR
by the US/Netherlands/UK IR Astronomical Satellite (IRAS). The smaller
detector footprint of the MSX detector arrays, as compared to IRAS, avoids
the confusion problems encountered by the IRAS survey in some areas of the
LMC, and allows a more precise determination of source position, which is of
great importance when attempting to compare these mid-IR observations with
the 2MASS near-IR observations. The MSX A band [6.8 -- 10.8 $\mu $m] is also
more sensitive than the IRAS 12 $\mu $m band [7 -- 15 $\mu $m], so therefore 
we can catalog 1806 sources to a limiting magnitude $\sim $7.5. 
The utility of mid-IR observations in identifying 
post-main-sequence stars in the LMC was demonstrated by IRAS, which, 
with a detection limit of magnitude $\sim 6$, detected several hundred extreme
evolved stars (such as OH/IR stars) in the LMC. At the distance modulus of 
the LMC, only the extremely
red objects at the brighter end of the mid-IR magnitude scale could be
detected by IRAS. By extending the LMC mid-IR survey nearly two magnitudes
fainter we expect to see some of the less extreme members of the LMC giant
branch populations and other objects. 2MASS observations, extending to $%
J=15.8$ mag at a signal-to-noise ratio (SNR) of 10, have detected and
yielded information on first-ascent red giant branch, asymptotic giant
branch (AGB), and obscured AGB populations of the LMC (Nikolaev \& Weinberg
2000). These red populations have also been inventoried and analyzed using
the DENIS point source catalogs by Cioni et al.~(2000) and other papers by
the DENIS team. By combining the near-IR and mid-IR data, we greatly expand
the color baseline and acquire additional insight into the AGB and other red
stars in the LMC, as well as discovering objects with unusual IR excesses.
In this paper we present the results of our merging of the MSX and 2MASS LMC
datasets and discuss the nature of the various IR-bright sources in that
galaxy, many of which are previously unidentified. This study provides a
foundation for future studies of the LMC by the {\sl SIRTF\/} and {\sl SOFIA}
missions.

\section{Observations}

\subsection{The MSX LMC Observations}

Details on the MSX spacecraft and experiments can be found in Price et
al.~(2001).  The IR telescope on MSX, known as SPIRIT III, is a 35-cm
clear aperture off-axis telescope with five line-scanned IR focal-plane
arrays and an aperture-shared interferometer. The entire system was cooled
by a single solid H$_{2}$ cryostat. The Si:As BiB arrays had eight columns
of detectors, each consisting of 192 rows of $18{\farcs3}$ square pixels. A
block consisting of half the columns in each array was offset by half a
pixel, providing Nyquist sampling in the cross-scan direction. The system
parameters are presented in Table 1.

In addition to the primary Galactic Plane survey experiments (Price et al.~2001),
MSX experiments also collected IR imaging radiometer data on selected
high-density regions, including the LMC and the SMC. The selected area scans
were executed at a slower scan rate than the Galactic Plane and IRAS Gap
long scans (0.05$^{\circ}$ s$^{-1}$ {\it vs.\/} 0.125$^{\circ}$ s$^{-1}$),
resulting in a slight increase in single scan sensitivity (cf. Table 1 in
this paper with Egan et al.~1999, Table 1). The sensitivity of band B, which
was designed for experiments looking into the Earth's atmosphere, is such
that no useful data were obtained for the LMC. The MSX LMC observations
covered 100 square degrees, centered on $\alpha =5^{h}21^{m}$, $\delta
=-68\arcdeg 45\arcmin$ (J2000). The full dataset includes six separate
data collection events, interleaved to provide a minimum of four redundant
passes over each area in the field.

The final data products for the LMC include FITS format images and a source
catalog. Both products are available through the NASA/IPAC IR Science
Archive (IRSA)\footnote{%
http://irsa.ipac.caltech.edu.}. The image data includes full resolution
image plates and a large-format full-resolution mosaic. Images were
constructed using the method described in Price et al. (2001), after
pointing refinement, dark current correction, and artifact mitigation have
been applied. The large-format mosaics of the galaxy are $1000\times 1000$
pixel FITS images in MSX bands A, C, D, and E, covering the full 100 square
degree area surveyed by MSX. The mosaics use 36\arcsec\ sampling, with the
resolution of the images degraded to 72\arcsec. The full resolution image
plates consist of 25 $1200\times 1200$ pixel, or $2\arcdeg\times 2\arcdeg$,
FITS images, in each band. These are sampled on 6\arcsec\ centers. The actual
resolution of these data is 20\arcsec.  The image products combine all of
the available scans over a given area. For four scans, this results in a
sensitivity increase of a factor of two in SNR over the single scan survey
data.

A preliminary version of the LMC point source catalog, produced to support
the {\sl WIRE\/} mission PV phase, was released in 1999 through IRSA. It is
this catalog on which the analysis presented in this paper is based. The
point source processing was identical to that described for the Galactic
Plane and also the Areas Missed By IRAS survey MSX Point Source Catalogs,
described by Egan et al.~(1999). Sources were extracted from individual scan
legs, and redundant scans were used to confirm point source detections and
increase the reliability of the catalog. In Figure 1 we show the number of
sources as a function of magnitude ($d{\log }N/dM$ {\it vs.\/} mag).
Here, we use the conversion to zero magnitude for a flux density of 58.49 Jy
in band A, 26.51 Jy in band C, 18.29 Jy in band D, and 8.75 Jy in band E
(Cohen, Hammersley \& Egan 2000). Version 1 of the LMC point source catalog
contains 1806 sources. Due to the fact that MSX band A was approximately 10
times more sensitive than the other bands, most sources are detected only in
this band. Given the increased sensitivity due to coaddition of the image
data, sources up to a factor of two fainter will be detectable in the images
for all bands. Version 2 of the MSX Point Source Catalog (PSC), which will
contain point source extractions from all of the MSX IR data collections,
including the LMC, will extract point sources from the co-added images in
bands C, D, and E, using the prior knowledge of the band A source positions.
This will yield a much higher fraction of sources with multi-color MSX
mid-IR data. This catalog is currently in preparation.

We have cross-referenced the MSX LMC sources with the catalog produced by
Schwering (1989) and Schwering \& Israel (1990) from the IRAS Additional
Observation (AO) program scans. These scans covered a smaller area than the MSX
survey (approximately 70 square degrees).  The goal of the IRAS AO survey
catalog was to provide a deep (magnitude $\sim $6, or $\sim $150--200 mJy,
at 12 $\mu $m) IR survey of the LMC. The scans yielded a catalog containing
1823 sources (802 of which were not contained in the IRAS Point Source
Catalog). Of the Schwering \& Israel sources, 1575\ have reported 12 $\mu $m
fluxes and 1578 have 25 $\mu $m measurements. We found that only $\sim $47\%
of the IRAS sources correspond with MSX sources. Overlaying the IRAS sources
on the A-band mosaic of the LMC, we find that only $\sim $71\% of the
cataloged 12 $\mu$m sources correspond to emission detected at 8.3 $\mu$m
by MSX. 

The sources not seen in the MSX 8.3 $\mu$m observations tend to
be associated with Schwering \& Israel sources having uncertain flux values
at 12 $\mu$m (denoted by a colon in their tabular
data).  More than half (56\%) of these sources do not have MSX counterparts
(either point sources, extended emission, or multiple MSX sources) {\it vs.\/}
only $\sim 30\%$ of those with IRAS 12 $\mu$m flux values with which Schwering
\& Israel were more confident.   Plotting the Schwering \& Israel
sources on an IRAS $F_{\nu}(60\ \mu {\rm m})/F_{\nu}(25\ \mu {\rm m})$ 
{\it vs.\/} $F_{\nu}(12\ \mu {\rm m})$ color-magnitude diagram, the sources 
not seen in the MSX 8.3 $\mu$m data are found to be the faintest 
($F_{\nu}(12\ \mu {\rm m})\leq 0.3$ Jy) and
reddest ($F_{\nu}(60\ \mu {\rm m})/F_{\nu}(25\ \mu {\rm m})>3$) sources on the 
diagram. Of the total 801 Schwering \& Israel sources in this quadrant of the
color-magnitude diagram, 610 (254 with uncertain 12 $\mu$m flux listings)
are not seen at 8.3 $\mu$m. While some of the MSX non-detections
may indicate spurious sources in the Schwering \& Israel catalog, it
appears that most are likely cases of cool, far-IR emission sources
which are below the detection limit in the MSX 8.3 $\mu$m band.

\subsection{The 2MASS LMC Observations}

2MASS has surveyed the entire sky in the near-IR from Mt.~Hopkins, AZ, and
the Cerro Tololo InterAmerican Observatory (CTIO). The cameras on the two
automated 1.3-m telescopes observed simultaneously in three channels, $J$
(1.25 $\mu $m), $H$ (1.65 $\mu $m), and $K_{s}$ (2.17 $\mu $m), using $%
256\times 256$ HgCdTe detector arrays, sampling the sky in $6^{\circ }\times
8^{\prime }$ strips. Coadded images, with 7.8-s total integration, are
produced from six dithered frames, after rebinning to 1$^{\prime \prime }$
pixels. The 2MASS Production Processing System provides final Atlas Images
and source extractions with precise photometric calibration and astrometric
positions, with 10$\sigma $ sensitivities of 15.8 mag at $J$, 15.1 at $H$,
and 14.3 at $K_{s}$.

The LMC was observed as part of routine nightly southern operations in 1998
-- 2000. Production processing resulted in nearly 7.1 million source
extractions for the LMC in a working point source database, which also
includes image artifacts, such as filter glints and diffraction spikes from
bright stars, as well as confused objects, detection upper limits, and
multiple source apparitions due to scan overlaps. Nikolaev \& Weinberg
(2000) discuss the results of analysis on the working point source database
for the LMC. Subsequently, artifacts, which can be well-characterized and
identified, were eliminated, and duplicate sources were removed, as part of
the catalog generation process for the 2MASS Second Incremental Data Release
(2IDR) in 2000 March. The 2IDR contains $\sim $47\% of the sky, including
most of the observed field of the LMC, which is comprised of 1,399,637 point
sources and 12311 extended sources.

Two relatively small sections of the galaxy (15.6 and 14.8 square degrees,
respectively) were not part of the 2IDR. The first section, spanning
approximately right ascension $\alpha=4^h39^m$ to $4^h28^m$ and declination $%
\delta=-72$\arcdeg\ to $-$66\arcdeg\ (J2000), coincidentally, had the
highest source density. The second, covering $\alpha=4^h00^m$ to $4^h05^m$ and $%
\delta=-69$\arcdeg\ to $-$64\arcdeg\ (J2000), is somewhat less dense. Data
for these sections were drawn from the working point source database, and
therefore may be contaminated by artifacts and confused and duplicate
sources. The combined released and unreleased datasets form the basis of our 
near-IR/mid-IR point
source cross-correlation and subsequent analysis. We have also
cross-correlated the MSX PSC with the 2MASS extended source catalog (XSC),
drawn from both the 2IDR and the working database for the two strips of
unreleased data.

\section{Analysis}

\subsection{Correlating The Observations}

The MSX and 2MASS catalogs have been cross-referenced using a simple
positional correlation, requiring the position error goodness-of-fit
statistic, $\chi ^{2}$, to be less than $18.4$, where 
\[
\chi ^{2}\equiv \frac{(\Delta \alpha /2)^{2}}{(\sigma _{\alpha
,2MASS}^{2}+\sigma _{\alpha ,MSX}^{2})/2}+\frac{(\Delta \delta /2)^{2}}{%
(\sigma _{\delta ,2MASS}^{2}+\sigma _{\delta ,MSX}^{2})/2}.
\]
For a two-dimensional normal error distribution, we expect 99.99 of the
source matches to be found within this $\chi ^{2}$ limit. Figure 2 shows the
distribution of $\chi ^{2}$ values for matches between the MSX catalog and
the 2IDR. The figure also shows the cumulative fraction of source matches as
a function of $\chi ^{2}$, which for a truly normal distribution of errors
would yield 68.3\% of sources with $\chi ^{2}<2.3$, 95.4\% with $\chi
^{2}<6.17$, and 99.73 with $\chi ^{2}<11.18$. The distribution of the $\chi
^{2}$ statistic found for this dataset is quite close to this, giving us
confidence that most of the position matches are true. The driving
uncertainty in these matches is that of the MSX position, which is of order $%
1{\farcs}5$ in both in-scan ($\delta )$ and cross-scan ($\alpha $)
directions. (This is somewhat better than the quoted uncertainties in the
MSX PSC Version 1.2 [Egan et al.~1999], since the subsequent LMC pointing
refinement used 2MASS and MSX IR Astrometric Catalog [Egan \& Price 1996]
stars, which yielded more pointing update stars per square degree for the
LMC refinement {\it vs.\/} the Galactic Plane boresight pointing refinement,
and reduced the MSX LMC positional uncertainties.) 2MASS positional
uncertainties\footnote{%
See Cutri et al.~(2001), the 2MASS 2IDR Explanatory Supplement, at
http://www.ipac.caltech.edu/2mass/releases/second/doc/explsup.html.} are
generally on the order of $0{\farcs}2$. Given the high source density,
especially in the LMC bar, many MSX sources had multiple 2MASS matches with $%
\chi ^{2}<18.3$, though the percentage is smaller than was expected. In
these cases, we have retained the match which had the smaller positional
error, except in a dozen cases for which both 2MASS sources were within 1$%
\sigma $ (rms) of the MSX position. In these cases, the redder (in $J-K_{s}$%
) of the 2MASS sources was chosen to be the true match.

Of the 1806 MSX catalog sources, 1664 had positional matches with 2MASS
point sources. Of these, 1488 were in the area covered in the 2IDR, and 176
are in the unreleased regions. In Tables 2 and 3 we list the first 45
sources from both the released and unreleased 2MASS datasets, respectively.
(The full list of sources is electronically available from the CDS
database.) In both tables we give the 2MASS source positions and magnitudes,
the MSX A-band magnitude, and the source type based on our cross-correlated
MSX-2MASS colors (see below).

Of the 1664 matched sources, 1036 had unique 2MASS matches (i.e., only one
2MASS source fell within $\chi ^{2}<18.4$). Most of the multiply matched
sources were fairly clear-cut cases, with one match of the pair lying within 
$\chi ^{2}<2$ and the other with $\chi ^{2}>6$. 142 MSX sources remained
unmatched with 2MASS point sources and are listed in
Table 4. We have also run the MSX PSC against the SIMBAD database, with a
match radius $0{\fdg}003$ ($10{\farcs}8$). These cross-references have also
been included in Tables 2 -- 4.

\subsection{Colors of Cross-Identified Sources}

Most (1576 in number) of the cross-identified sources have measurements in
at least four bands, the 2MASS $J$, $H$, $K_{s}$ and the MSX A band. These
bands can be used to identify six color axes ($J-H$, $J-K_{s}$, $J-{\rm A}$, 
$H-K_{s}$, $H-{\rm A}$, $K_{s}-{\rm A}$). We have used a variant of a
``fuzzy clustering'' algorithm (see, e.g., H\"{o}ppner et al.~1999) for $N$%
-dimensional data to find clusterings of sources with similar color
characteristics. These clusters identify specific populations of celestial
objects for those populations that have unique colors. Processing first only
the 1488 matched 2IDR objects through the algorithm (since they are expected
to be far less contaminated by artifacts and multiple apparitions than is
true for the unreleased data), we find 19 distinct groups. Some of the
groups are subgroups of a parent type, e.g., the method distinguishes
between oxygen-rich AGB stars of spectral types both earlier and later than
subtype M6. We have combined these subgroups under their main group type and
come up with a total of 11 major categories.

We have assigned categories for all matched sources and include the source
types in Tables 2 -- 4. We note that 14\% (228) of the stars are not
assigned candidate types, either because they fall between the chosen
category boundaries (true for 160 sources) or are saturated in one or more
2MASS bands. 

Six-dimensional data is difficult to represent; additionally, it is not
clear that all of the colors yield completely independent information. To
simplify the analysis we have chosen the three axes which most clearly
illustrate the object type distinctions. Figure 3 is a $J-K_{s}$ {\it vs.\/} 
$K_{s}-$A color-color diagram, where the colors of the points represents the 
$H-K_{s}$ color. Photometric uncertainties (rms) in each color are also
shown. The $J-K_{s}$ {\it vs.\/} $K_{s}-$A color plane shows several obvious
groupings, while the $H-K_{s}$ color is able to further separate the other
groups. To identify the sources contained in the cluster-analysis
categories, we have examined the colors expected for the various source
types by the Wainscoat et al. (1992) ``SKY'' model of the Galaxy and
compared these to the observed colors. Absolute magnitudes in the MSX A band
were supplied for each of the 87 objects by M.~Cohen (priv.~comm.). \
Figure 4 is the model analog to Figure 3.

The 11 categories occupying the three-axis color-color diagram in Figure 3 and
the number of sources assigned to each group, are as follows:

\begin{itemize}
\item  {\bf I }- Early dwarfs. \ These stars occupy the region defined by $%
K_{s}-$A$<0.5$, $J-K_{s}<0.5$. There are 50 candidates in the sample.

\item  {\bf II} - Late dwarfs, defined by $K_{s}-$A$\leq 1.0 $, $0.5\leq
J-K_{s}<1.0$. Our sample contains 19 candidates.

\item  {\bf III} - Early giants (primarily G III and K III spectral types),
occupying the region of the diagram, $K_{s}-$A$<0.3$, $0.5\leq J-K_{s}<0.9$.
There are 251 objects in this category.

\item  {\bf IV} - Late giants. The M III\ types are contained in the region
bounded by $K_{s}-$A$<0.5$, $0.9\leq J-K_{s}<1.4$. There are 176 candidates.
\end{itemize}

Category V contains a mixture of several types of stars. From Wainscoat et
al.~(1992) we see that the colors of red supergiants (RSGs) and early 
oxygen-rich
and carbon-rich AGB stars are all quite similar. (By ``early'' we mean those
stars with relatively little mass loss or obscuration of the central star at
visible and near-IR wavelengths). To a lesser degree, we then use the $%
H-K_{s}$ color to distinguish between these objects, the RSGs
being the bluest objects among this group on this axis, and the carbon stars
being the reddest. However, we do not claim that these designations are
exact. In fact, even using all six color axes, some confusion still exists
between some of these sources. The designations below try to make the best
separation between subtypes in this category:

\begin{itemize}
\item  {\bf Va} - RSGs, enclosed by $1.0<J-K_{s}<1.4$, $%
0<H-K_{s}<0.5$, and $0.5<K_{s}-$A $<1.75$. We find 99 likely candidates.

\item  {\bf Vb} - Early oxygen-rich AGB stars, within the color ranges, $%
0.4<H-K_{s}<0.75$, $0.1<K_{s}-A<7.0$, and $\frac{(K_{s}-{\rm A})}{7}%
+1.0<J-K_{s}<1.5\left[ \frac{(K_{s}-{\rm A})}{7}+1.0\right] $. We find 61
candidates.

\item  {\bf Vc} - Early carbon-rich AGB stars, having colors within the
ranges, $0.5<H-K_{s}<1.25$, $0.2<K_{s}-A<1.5$, and $1.4<J-K_{s}<1.75$. Only
four stars appear to be early C-rich AGB stars.
\end{itemize}

Examining just the $J-K_{s}$ {\it vs.\/} $K_{s}-$A two-color diagram,
category VI appears to be a monolithic group. The six-dimensional analysis
of the observed colors and the SKY model reveal that it actually contains
two distinct types of objects: the ``late'' (that is, high mass-loss rate,
high obscuration) C-rich AGB stars (also known as ``infrared'' carbon stars;
Chan \& Kwok 1988) and their O-rich counterparts (generally OH/IR stars and H%
$_{2}$O maser sources). In this category, the mixing is not as thorough as
in category V, so the C-rich and O-rich sources are more easily separated.
However, we still expect some confusion in the overlap region of these
categories:

\begin{itemize}
\item  {\bf VIa} - C-rich late AGB stars, within $1.5\leq K_{s}-A<3.75$ and $%
0.8<H-K_{s}\leq 1.5$, yielding a total of 108 likely IR carbon stars.

\item  {\bf VIb} - O-rich late AGB stars (OH/IR stars), within the region
bounded by $J-K_{s}>3$, $K_{s}-A$ $\geq 3.75$, and $H-K_{s}>1.5$. There are
337 likely OH/IR stars.
\end{itemize}

Category VII represents objects with rather unusual colors. The near-IR
colors are blue, and on the $JHK_{s}$ color-color diagram they lie among the
normal stars. However, they show a very large excess emission component at
8.3 $\mu $m, indicative of dust associated with a stellar point source. The
SKY model colors and the SIMBAD identifications show that the associations
of the 2MASS sources and the MSX sources are indeed real. The objects in
this group tend to be either planetary nebulae (PNe), or early (O- and
B-type) stars associated with dust (i.e., HII regions, reflection nebulae,
LBVs, dusty B stars, etc.). To some degree we can separate the objects based
on the $H-K_{s}$ color, with the PNe being redder than the other objects.
However, these objects tend to be among the fainter of the 2MASS sources.
Therefore, confusion still exists in identifying these sources near the
assumed $H-K_{s}$ boundary:

\begin{itemize}
\item  {\bf VIIa} - PNe or PPNe, with colors $J-K_{s}<2$, $K_{s}-$A $\geq 2.5
$, and $0.75\leq H-K_{s}<2$, in accord with the range of colors seen in the
SKY model for blue and red PNe, and for the known PNe from the SIMBAD
identifications (see below). This yields a total of 76 candidate sources.

\item  {\bf VIIb} - HII regions, dusty reflection nebula-like objects,
likely associated with star forming regions and OB associations, with $%
J-K_{s}<\frac{(K_{s}-{\rm A})}{7}+1.0$, $K_{s}-$A $\geq 3.75$, and $%
H-K_{s}<0.75$. There are 254 candidates.
\end{itemize}

\subsection{Identification of Specific Objects}

As a check of the proposed object identifications resulting from the
cluster-analysis and model colors, we have compared the 2MASS/MSX proposed
spectral types to those contained in the SIMBAD database. 933 sources were
returned with some type of identification, as listed in Tables 2--4. The
relatively good agreement between the SIMBAD spectral types/object
identifications and our object type assignments gives us a high degree of
confidence in our method. As candidates become more IR-bright, they tend not
to have optical counterparts and are therefore less likely to have spectral
types. We discuss the detailed results of our cross-referencing here.

All of the stars in category I and almost two-thirds of category II have
identifications (mostly HD numbers) in the SIMBAD database. Category I
(early dwarfs)\ consists of B through F dwarfs (32 luminosity class V
types), with a few emission-line stars, rounded out by ten
subgiants, five A--F giants, and an A\ supergiant. Of the nineteen category
II sources (late dwarfs), seven do not have spectral types in SIMBAD, and 11
have only spectral types without a luminosity class (three M, four K, and
four G type stars). The only source in this category with both a spectral
type and luminosity class, HD 33060, is listed as G8/K0III, a late giant.

The candidate giants (categories III [early] and IV [late]) show similar
results. All but one source in category III have identifications in the
SIMBAD database. Half of these sources are identified as G or K giants. Most
of the remaining sources have only spectral types, primarily of G or K type,
although six are classified as M-type stars. In category IV, 147 sources are
listed in SIMBAD, and 95 of these have spectral type information. Of the 22
with both spectral types and luminosity classes, 19 are K giants. For stars
without luminosity class information, 18 are K-type and 54 are M-type stars.

Category V, the supergiants and low mass-loss-rate AGB stars, contains less
certain classifications. The supergiants are the least ambiguous, with 13 of
the 99 candidates of M type and luminosity class Ia or Iab, and one
classified as K5Iab. Only one of the four C-rich AGB candidates has a
spectral type. The 61 O-rich AGB candidates include 2 Mira variables, a
semi-regular variable, and 23 M-type stars, as well as two supergiants and a
C-rich Wolf-Rayet star.

Category VI, which consists of the more heavily obscured AGB stars, have
only a few reliable spectral-type identifications. Of the 337 O-rich
candidates, only 55 appear in SIMBAD, and none of these have spectral types.
One is classified as a PN, and a few are noted to be variable stars, but the
majority (34) are simply listed as ``IR sources'' (mostly from IRAS PSC
listings). We find a few spectral types among the C-rich candidates, with
five of the 17 stars with SIMBAD references classified as carbon stars. The
remainder are, again, simply listed as variable stars, or one, as member of
a cluster.

In category VII, 40 of the 76 of the PN candidates have SIMBAD
identifications. Seventeen are listed simply as ``IR source'', while eight
are ``emission object'' or ``emission star.'' Only one is actually listed as
a PN, and the remaining types reflect a wide variety of unusual objects
(e.g., nebulae, line sources, UV sources, radio sources). Our HII region
candidates have similar sorts of SIMBAD identifications for 75 of the 254
sources. ``IR\ sources'' account for 28 of these objects, while eighteen are
listed as HII\ regions, nebulae, or clusters. Fifteen are ``emission
objects,'' and the remainder are stars or radio sources.

\subsection{LMC and Foreground Stellar Populations}

To properly study stellar populations in the LMC using our 2MASS/MSX
cross-correlation, we must first separate the foreground Milky Way stars
from the actual LMC sources. To do so, we can examine the $K_{s}$ {\it vs.\/}
$K_{s}-$A color-magnitude diagram shown in Figure 5. The dashed line is the
limit represented by the depth of the MSX survey in band A, approximately
magnitude 8, and the solid line shows the magnitude limit for $K_{s}$ at
SNR=10. The points have been color-coded to reflect their $J-K_{s}$ colors,
so they can be more easily compared to the color-color diagram in Figure 3.
In the color-magnitude diagram, the sources segregate into four distinct
groups. From brightest to faintest at $K_{s}$, we find: (1) A
``crescent-shaped'' group with $K_{s}<8$, with $K_{s}-$A $\sim 0$, in the
range $-$0.75 to 0.5. For this group, $J-K_{s}<1.5$, and its members
encompass the elements of categories I--IV from the color-color diagram. (2)
The knot of sources clustered at $K_{s}\simeq 8$, with $K_{s}-$A $\sim 1$,
and $1.5<J-K_{s}<2.5$, which corresponds primarily to category V, although
some overlap exists with the M dwarfs. (3) Those sources with $K_{s}>9$, $%
K_{s}-$A$>2$, and $J-K_{s}>2.5$, encompassing mainly the members of category
VI. (4) Finally, the PNe and HII regions of category VII can be seen as
those sources generally with $K_{s}>12$, $K_{s}-$A$>2.5$, and $H-K_{s}<2$.

The apparent $K_{s}$ magnitude of a source in our sample will, of course,
depend on the absolute $K_{s}$ magnitude, $M_{K_{s}}$, of the source and its
distance modulus, $\mu $. Either the object is in the LMC, with $\mu \simeq
18.5$ (e.g., van Leeuwen et al.~1997, Gratton 1998, Koen \& Laney 1998), or
the object is in the Milky Way, with $\mu $ likely between 10 and 14 ($\sim $%
1 kpc and $\sim $6 kpc, respectively). For convenience, we shall assume $\mu
=12$ ($\sim $2.5 kpc) as typical for foreground stars in the direction of
the LMC. The SKY model contains the absolute magnitudes of each of the 87
source types which describe point sources in the Milky Way. We have modeled
Figure 5 using these source types, for $\mu =12$ in Figure 6, and $\mu =18.5$
in Figure 7, ignoring differences between $K$ and $K_{s}$.

Figure 6 indicates that a distance modulus of 12 represents the M giants in
the diagram fairly well. However, we know from the color-color diagram that
a number of the stars are likely G and K giants and main sequence stars, and
that a few M dwarfs are in the sample. The figure implies that the more
likely distance modulus for any G and K candidate giants is $\mu \sim 8$ ($%
\sim 1$ kpc), and that the main sequence stars we find are likely closer
than 1 kpc, and probably more like 100 pc. From Figure 7 we see that T Tauri
stars in the LMC are simply too faint to have been detected by MSX. This is
also true for the main sequence and giant branch stars. The brightest M
giants would be about magnitude 9 at 8.3 $\mu $m if they were located in the
LMC, one magnitude fainter than the MSX\ band A survey limit (however, these 
might possibly be detected in the co-added MSX image data). Most of the LMC M
giants, while easily detectable by the 2MASS survey at $K_{s}\sim 13$, would
have magnitude $\sim $13 at 8.3 $\mu $m, undetectable by MSX (even in the
co-added data). For the reddest M dwarfs in the sample, it appears that the
furthest away they could be and still have been detected by MSX is $\sim $25
pc.

Figures 6 and 7 reveal that the reddest candidate objects, i.e., the heavily
enshrouded AGB stars and the RSGs, are intrinsically so bright at 
$K_{s}$ that most in our sample must have $\mu >15$, making them most surely
extragalactic. The PN and most HII region candidates are generally fainter
at $K_{s}$, but given their colors, are also most likely extragalactic. It
is possible that some of the earliest AGB star candidates (both C-rich and
O-rich) could be extreme members of the Milky Way. 
Figure 7 suggests that the early carbon stars in the LMC would be about a
magnitude too faint to have been detected in MSX band A (assuming they have
the same limiting absolute magnitude in the LMC and the Milky Way). This
likely explains the fact that we identify only five early AGB stars.

\section{Discussion}

\subsection{Comments on Assigning Object Types}

From the color-color diagrams the most difficult assignments are for the
C-rich and O-rich AGB stars. These tend to blend together in color,
especially for those with little dust emission. An added complication is the
fact that the difference in metallicity between the Milky Way and the LMC
makes it unlikely that the use of the Wainscoat et al.~(1992) model yields
entirely accurate results for these stars. However, the IR carbon stars
appear to agree in color with the Wainscoat et al., which is corroborated by
the SIMBAD cross-referencing. Thus, no {\it a priori\/} reason exists to
believe that the colors for the low mass-loss rate (visual) carbon stars are
incorrect.

We identify 61 early O-rich AGB stars in the sample, and only 4 (or, $\sim $%
6\%) early C-rich AGB stars. Among the dusty AGB stars, we find 337
OH/IR-type stars, and 108 (or, 24\%) IR carbon stars. Groenewegen \& de Jong
(1993) indicate that an approximately even number of O-rich and C-rich AGB
stars are expected in the LMC, which contrasts with the 10--13\% expected to
be C-rich in the Milky Way (Hacking et al. 1985; Thronson et al. 1987). The
lower metallicity of the LMC should result in an increased fraction of
carbon stars. We see a lower percentage of C-rich stars than Groenewegen \&
de Jong expect, but this may be due to the fact that the O-rich population
(OH/IR stars) is intrinsically brighter than their C-rich counterparts.
Thus, more O-rich stars are observed by MSX, given its band A limiting
magnitude.

One of the surprises of this study was the existence of the grouping of
sources with $K_{s}-$A $>6$ and $J-K_{s}\sim 0.5$, the category VII objects.
We had found a fairly large number of stars with stellar near-IR colors, but
with large mid-IR excess. Furthermore, the group segregates into two near-IR
subgroups, a very blue one and a redder one. 
The Wainscoat et al. (1992) colors indicated that the redder $H-K_{s}$ (and,
to a lesser degree, in $J-K_{s}$) group correspond most closely to the ``red
planetary'' colors in the model. Of the few PNe identified in SIMBAD, these
tend to fall in or near the colors defined for category VIIa.

Many of the stars we identify as PNe have spectral types indicating peculiar
emission-line stars/objects, which is consistent with the PNe central stars.
These identifications are the least likely in the sample, however, given the
overlap with the HII region sources. It may be that a number of these are
HII regions with more reddening than the others, rather than PNe.

To more adequately determine source types in Tables 2--4, it is clear that 
follow-up observations need to be made for many or all of the sources,
including further imaging, but, particularly, spectroscopy. A preliminary set
of optical spectroscopic observations by one of us (SVD) of a few objects in
each of the categories (except HII regions) shows that our classification
scheme has merit, particularly for the RSGs, AGB stars, and PNe. However,
ground- and space-based IR spectroscopic and imaging observations are
essential, especially for the more dust-obscured sources, such as the OH/IR
and IR carbon stars.

\subsection{Spatial Distribution of Object Types}

Figures 8a and 8b show the location of objects in the 11 candidate
categories distributed over the MSX 8.3 $\mu$m image mosaic of the LMC.
These figures can be quite useful in analyzing the validity of the
identifications we have made on the basis of source color, and may also be
of use in understanding star formation and evolution in LMC. Figure 8a shows
the sources of luminosity class V and III, the most numerous being the
giants. We have previously noted that the main sequence stars and giants in
the sample must be foreground (Milky Way) objects, which we would not expect
to be correlated with any structures in the LMC. Indeed, we see that these
stars are distributed fairly randomly across the image, with no apparent
clustering or association with any of the mid-IR emission in the LMC.

Figure 8b shows the distribution of the remaining sources, which we argue
most likely belong to the LMC. The primary population are the extreme AGB
stars. These are expected to be fairly massive (up to 8 $M_{\odot }$) and
probably have ages of $\lesssim $1 Gyr. 
These sources are fairly widespread over the face of the LMC, with a
concentration along the LMC bar. The next most populous groups are younger
objects, i.e., large population of supergiants (99 in number, ages $\lesssim 
$0.5 Gyr), and a somewhat larger population (254 sources) of HII regions or
reflection nebulae (expected to contain OB associations, ages $\lesssim $%
10--30 Myr). The mid-IR bright RSGs are confined to the central
portions of the LMC, with two major clusterings. The first of these lies to
the immediate west of the 30 Doradus complex, and northeast of the LMC bar,
centered near $\alpha =5^{h}30^{m}$, $\delta =-69^{\circ }$ (note that SN
1987A is at the eastern edge of this cluster, at $\alpha =5^{h}35^{m}28^{s}$%
, $\delta =-69^{\circ }16^{\prime }13^{\prime \prime }$, J2000). The second
large cluster of supergiants is almost due north of the first, centered near 
$\alpha =5^{h}32^{m}$, $\delta =-67^{\circ }$.

The largest concentrations of HII\ region candidates (category VIIb sources)
trace the LMC bar, and many of the remaining HII\ regions are associated
with other knots of mid-IR emission. There are also a large number of
apparently randomly distributed category VIIb sources. An interesting
comparison between the distribution of RSG and HII region
candidates seen in the mid-IR and the distribution of the youngest LMC stars
(bright main sequence dwarfs, blue loop stars and supergiants) identified by
the DENIS near-IR survey (Cioni et al. 2000) can also be made. The contours
defined in Cioni et al., their Figure 2, which trace the star count density
of young ($<0.5$ Gyr) stars, are qualitatively very similar to the
distribution of RSGs and HII regions in Figure 8b, including the
enhancements in counts seen around the 30 Doradus and N11 regions. The
distribution of HII regions tends (unsurprisingly) to trace the mid-IR
extended emission. It is apparent that the current high-mass star formation
in the LMC is strongly concentrated in or near the bar, although knots of 
mid-IR emission trace other regions of high-mass star formation throughout 
the LMC.

The stars we identified as ``early'' C-rich AGB stars are all found in the
northeast quadrant of the LMC, while the ``late'' C-rich AGB stars are found
in two primary areas, one just south of the LMC bar, and another in the
northwest quadrant of the galaxy, around the N11 complex. The O-rich AGB
stars are more widely distributed, but it appears that their distribution
traces the areas of bright mid-IR emission.

The distribution of candidate PNe are fairly widespread over the face of the
LMC, although there is some association with knots of mid-IR emission along
the bar. The likely PNe are more closely associated with the central
portions of the LMC than are the HII region members of category VII. There
is a cluster of $\sim$6 PNe in the 30 Doradus complex, as well as another
cluster of three sources about $0{\fdg}5$ to the northwest. Another group of
PNe runs through the middle of the bar.

\subsection{Unmatched MSX Sources}

There are 142 sources in the MSX PSC which did not have matches in the 2MASS
PSC. A comparison of the magnitude distribution of the matched and unmatched
MSX sources is shown in Figure 9. We have visually examined the 2MASS and
MSX image data, and find that lack of matches can generally be explained by
four reasons: 1) a source is too bright (i.e., saturated) in 2MASS, 2) a
source is too faint in 2MASS, 3) a possible positional mismatch exists, 4)
the 2MASS source is extended, or 5) something appears in the 2MASS ($K_{s}$)
image, but was not detected, for whatever reason.  The 2MASS point source
processing for the 2IDR could not extract extremely bright sources, but 
instead, placeholders for
these stars were put in the 2IDR PSC (with values $-$99.999 for magnitudes).

Five of the unmatched sources were far too bright for 2MASS. We have
identified these stars, matched them with SIMBAD, and extracted their
(inaccurate) magnitudes from the 2MASS working database: HD 31907 (M1III), $%
K_s \sim 3.2$; HD 271114 (F0), $K_s \sim 2.7$; WOH G 622 (M-type), $K_s \sim
3.1$; HD 271928 (M4), $K_s \sim 3.1$; and, HD 272138 (M2), $K_s \sim 2.9$.
An additional five MSX sources have nearby 2MASS counterparts, and we
suspect that the MSX pointing correction may have been locally inadequate in
these cases, with an optimistic quoted positional uncertainty, resulting in
a failure to make the proper match. Nineteen MSX sources appear correlated
with relatively faint 2MASS sources which appear extended. Five of these
faint sources are in fields with general nebulosity, i.e., brighter extended
sources in the 2MASS XSC which are also correlated with MSX point sources;
in fact, we can account for one ``unmatched'' 2MASS PSC source, since it is
actually an extended source, matched with MSX (see below).

The overwhelming majority of unmatched sources do not appear on 2MASS
images, likely because they are too faint for 2MASS (i.e., so red) that the $%
JHK_{s}$ magnitudes are below the detection limits.  However, a fraction of
the sources are probably spurious sources in the MSX PSC. Figure 9 shows
that most of the unmatched sources are near the MSX detection threshold. One
such spurious source is obvious in the case of MSX source no.~187, which
inadvertantly survived the catalog compilation with a bad SNR calculation.
Examination of the MSX image data indicates that 37 of
these objects are unambiguous point sources in the co-added image, six appear
to be multiple sources, and thus may have incorrect position information in
the catalog, 29 are somewhat non-point-like, and associated with extended
emission, 35 are at very low SNR on the image plate, but
probably are real, and 30 have no identifiable counterpart on the image
data and are likely spurious catalog sources.

\subsection{MSX Sources without Category}

Among the 228 MSX sources in Tables 2 and 3 without assignment to one of the
11 object categories, 85 of these have spectral type identification from
SIMBAD. 68 (or 30\%) of the uncategorized sources are saturated in the 2MASS
bands, but most of these have SIMBAD identifications consistent with what
are most likely foreground giant and dwarf stars. For the rest of the
uncategorized sources, 31 (or 14\%) are very red and relatively faint as
seen by 2MASS, 9 (4\%) are rather blue and faint, and 19 (8\%) are blended
or confused, such that the 2MASS and MSX colors are not representative of
the individual objects, and three are clearly nebular (with an associated
2MASS point source). The blended objects either are of similar color, e.g.,
in a compact star cluster, but are just too confused, or, in several cases,
are a pair of confused stars of very different colors. Among the nebular objects
is a known HII region (LI-LMC 102), a source associated with N 105A, and a
source associated with the emission object LHA 120-S 24. Among the faint,
red objects is a known LMC PN (IRAS 04515$-$6710), and we suspect that others
are in this category. We also suspect that many of the faint, blue sources
are (compact) HII regions, based on their colors.

Yet, most (97, or 42\%) appear stellar and have colors and SIMBAD
identifications (for some) that indicate that these are likely AGB or RSG
stars with a range of luminosities. Known sources in this group are HD
269953 (a LMC G0Ia star), WOH G 64 (a RSG, a known H$_2$O maser source), and
SHV 0510360-693335 (a long-period variable, possible carbon star). Many are
quite red, and, therefore, possibly dusty. An extreme example is the
unusually red MSX source no.~811, with $J-K_s=5.09$, $H-K_s=2.02$, and $K_s-$%
A=3.14. The fact that these stars did not fall directly into our object
categories may indicate that the Wainscoat et al.~(1992) colors for Galactic
objects do not fit well the colors for these specific LMC sources. Along
with the many sources in this study that fall into our categories, these
uncategorized objects also deserve follow-up IR imaging and spectroscopic
observations.

\subsection{Extended Sources in 2MASS}

There are 131 MSX sources matched with 2MASS extended sources across the
LMC. Although 2MASS detects extended sources over the entire 100
square-degree field, the central region of the galaxy will have the
preponderance of matched extended sources, since most of the MSX structure
is concentrated there. The matched sources are generally faint. The majority
($\sim$57\%) appear nebular, while a large number ($\sim$31\%) appear
point-like, and many of these are fairly red. About 12\% of the extended
sources are galaxies behind the LMC. 
Most of the extended sources fall in the same color regime as the candidate
PNe and HII regions, and a few are near the area on the diagram for the AGB
stars. The MSX footprint is large enough that many sources in the LMC that
are extended, as seen by 2MASS, would be considered ``point-like'' and would
fall into the MSX PSC. All of the matched extended sources are associated
with matched MSX/2MASS point sources. This occurs, since many 2MASS extended
sources contain associated point sources, such as ionizing stars within the
nebulae, in the case of the candidate HII regions or reflection nebulae, or
central stars within candidate PNe. Also, a few of the candidate AGB stars
must show associated extended emission, e.g., circumstellar nebulae, as seen
by 2MASS, which, together, are also bright in the MSX A band.

\subsection{Future Directions}

This work presents a classification scheme for resolved objects in external
galaxies where near-IR and mid-IR colors are available. In addition to
future large ground-based IR observing programs, the next opportunity for
large-scale application of this work would be to undertake a survey of the
LMC with the {\sl SIRTF\/} IR Array Camera (IRAC) instrument. Like MSX, IRAC
has an 8 $\mu $m capability. Its projected sensitivity is approximately
magnitude 12 in this band in the moderate integration mode. Combining an
IRAC LMC survey with the 2MASS point source catalog would yield a much more
complete population study of the LMC. The limits of such a survey are shown
on the magnitude-color diagrams in Figures 5 and 7.

The magnitude=7.5 limit of the MSX A-band survey was too high to detect the
majority of AGB stars with low-to-moderate mass loss. We see from Figure 7
that a {\sl SIRTF}/2MASS survey will detect the entire LMC AGB population,
as well as the fainter end of the PN population, assuming the Wainscoat et
al.~(1992) absolute magnitudes are similar to what the same stellar types
would have under the metallicity conditions in the LMC.

Based on the IRAC performance provided by Hora, Fazio, \& Willner (2000), an
IRAC survey would detect $m_{[8\ \mu {\rm m}]}=15.1$ sources at SNR=5 with
the 30-s integration mode (A.~Omont \& G.~Gilmore, priv. comm.). From the
MSX 8.3 $\mu $m source counts in Figure 1, confusion is estimated to set in
at $m_{[8\ \mu {\rm m}]}\sim 16$. Indeed, confusion is relatively minimal
in the densest regions of the LMC at a $K_{s}$ magnitude limit of 14.7. For
2MASS, SNR$\simeq $5 occurs at $K_{s}=15.1$. Thus, an IRAC [8 $\mu $m] LMC
survey is well-matched to the 2MASS survey of the LMC. Again, from Figure 1,
more than a million LMC sources would be matched with $J$, $H$, $K_{s}$, and
the IRAC colors. 2MASS also conducted a ``6X'' survey (six times the normal
survey integration time per scan) of the LMC, increasing the survey limit by
about 1--1.5 magnitudes. We recommend such a {\sl SIRTF\/} survey be done as
one of the major Guest Observer projects.

\section{Conclusions}

MSX and 2MASS have both observed the LMC as part of their operations in the
mid- and near-IR, respectively. We have performed a cross-correlation of the
source catalogs from the two IR surveys and have found 1664 point source and
131 extended source matches. It is clear that the combination of near-IR and
mid-IR colors can lift some of the degeneracy in the IR color-color diagram
for sources in the LMC. For instance, in the 2MASS-only color plane, HII
regions and PNe are indistinguishable from normal stars. With the MSX A-band
as another axis to the color plane, such nebulae are much easier to identify
and distinguish.

Based on the combined 2MASS/MSX colors, using the IR point source model by
Wainscoat et al.~(1992), and obtaining source names and spectral types from
the SIMBAD database, when available, we have identified 11 categories of
stellar populations and red nebulae, including main sequence stars, giant
stars, RSGs, C- and O-rich AGB stars, PNe, H II regions, and other dusty
objects likely associated with early-type stars. 731 of these sources were
previously unidentified. Object detection is limited by the MSX 8.3 $\mu$m
(band A) sensitivity.

The spatial distribution of these IR sources can also add to our
understanding of star formation and stellar evolution in the LMC, and of the
evolution of the LMC as a member of an interacting system with the Milky
Way, as well as a template for low-metallicity systems at higher redshift.
Our compilation of MSX and 2MASS photometry and astrometry for these sources
will provide a sample of objects for future observations and analysis by
ground-based and space-based observatories, such as {\sl SIRTF\/} and {\sl %
SOFIA}.

\acknowledgments
This research has made use of the SIMBAD database, operated at CDS, Strasbourg, 
France, and of NASA's Astrophysics Data System Abstract Service. This publication 
made use of data products from the Two Micron All Sky Survey, which is a joint 
project of the University of Massachusetts and the Infrared Processing and 
Analysis Center/California Institute of Technology, funded by the National 
Aeronautics and Space Administration and the National Science Foundation. 
We thank Martin Cohen for determining magnitudes in MSX bands for the 87 
Wainscoat object types, and also thank Sean Carey, Don Mizuno and Tom Kuchar, 
who assembled the MSX Image data.

\begin{deluxetable}{ccccccc}
\def\phmm{\phm{$-$}}
\tablenum{1}
\tablecolumns{7}
\tablewidth{5.0in}
\tablecaption{SPIRIT III Spectral Bands\label{Table 1}}
\tablehead{\colhead{Band} & \colhead{No.} & \colhead{Isophotal}
& \colhead{50\%} & \colhead{Isophotal} & \colhead{Abs.} &
\colhead{LMC Survey} \\
\colhead{} & \colhead{Active} & \colhead{$\lambda$($\mu m$)} & \colhead{Peak}
& \colhead{BW ($\mu m$)} & \colhead{Photom.} & \colhead{Sens.} \\
\colhead{} & \colhead{Cols.} & \colhead{} & \colhead{Intensity} & \colhead{}
& \colhead{Accuracy} & \colhead{(Jy)}}
\startdata
A & 8 & 8.28 & 6.8--10.8 & 3.36 & 5\% & 0.045 \\
B$_{1}$ & 2 & 4.29 & 4.22--4.36 & 0.104 & 9\% & 10 \\
B$_{2}$ & 2 & 4.35 & 4.24--4.45 & 0.179 & 9\% & 6 \\
C & 4 & 12.13 & 11.1--13.2 & 1.72 & 3\% & 0.6 \\
D & 4 & 14.65 & 13.5--15.9 & 2.23 & 4\% & 0.5 \\
E & 2 & 21.34 & 18.2--25.1 & 6.24 & 6\% & 1.1 \\
\enddata
\end{deluxetable}


\begin{deluxetable}{cccrrrcccl}
\rotate
\tabletypesize{\small}
\def\phmm{\phm{$-$}}
\tablenum{2}
\tablecolumns{10}
\tablewidth{8.75in}
\tablecaption{MSX Sources in Common with the 2MASS 2IDR\label{Table 2}}
\tablehead{\colhead{MSX \#} & \colhead{RA (J2000)\tablenotemark{a}} & 
\colhead{Dec (J2000)\tablenotemark{a}} & 
\colhead{$J$} & \colhead{$H$} & \colhead{$K_{\rm s}$} & \colhead{A (8.3 $\mu$m)} &
\colhead{IR Type} & \colhead{SIMBAD Type} & \colhead{SIMBAD Name} }
\startdata
 1 & 77.5366 & $-$64.3182 & \nodata & \nodata & \nodata & $-$0.58 && M8IIIe & Mi*$|$V* U Dor \\
 2 & 75.9857 & $-$65.0121 & \nodata & \nodata & \nodata & \phmm{2.46} && M4III & *$|$HD 33213 \\
 3 & 76.1875 & $-$64.5457 & 6.00 & 5.46 & 5.34 & \phmm{5.45} & A-K III & K0IIICN. & *$|$HD 33293 \\
 4 & 76.6619 & $-$65.3762 & 5.89 & \nodata & 5.12 & \phmm{5.33} & A-K III & K2III/IV & *$|$HD 33616 \\
 5 & 77.4312 & $-$64.3862 & 6.75 & 6.09 & 5.92 & \phmm{5.99} & A-K III & K3III & *$|$HD 34016 \\
 6 & 78.0128 & $-$65.1757 & 6.21 & 5.96 & 5.92 & \phmm{6.17} & MS(V) & F5V & *$|$HD 34349 \\
 7 & 77.4316 & $-$65.3665 & 8.84 & 8.07 & 7.71 & \phmm{6.23} & RSG && IR$|$IRAS 05095$-$6525 \\
 8 & 77.5507 & $-$65.3261 & 15.82 & 13.55 & 11.64 & \phmm{6.51} & OH/IR \\
 9 & 77.2655 & $-$63.9412 & 7.24 & 6.58 & 6.47 & \phmm{6.59} & A-K III & K2III & *$|$HD 33895 \\
10 & 77.5185 & $-$65.0441 & 6.94 & 6.37 & 6.20 & \phmm{6.49} & A-K III & K1III & *$|$HD 34074 \\
11 & 78.1615 & $-$64.2038 & 14.08 & 12.10 & 10.52 & \phmm{6.73} & OH/IR \\
12 & 78.0665 & $-$65.3120 & 16.49 & 13.80 & 11.89 & \phmm{7.12} & OH/IR \\
13 & 78.1952 & $-$64.2365 & 7.38 & 6.64 & 6.45 & \phmm{6.71} & MIII && *$|$CPD$-$64 428 \\
14 & 76.9128 & $-$64.5418 & 7.94 & 7.40 & 7.27 & \phmm{7.45} & A-K III && *$|$HD 271019 \\
15 & 77.8264 & $-$65.4357 & 8.70 & 7.91 & 7.66 & \phmm{7.44} & MIII & M & V*$|$SV* HV 2886 \\
16 & 77.7092 & $-$64.2588 & 7.63 & 7.13 & 6.99 & \phmm{6.97} & A-K III & ? & *$|$CD-64 178 \\
18 & 77.1345 & $-$65.1374 & 12.10 & 11.03 & 10.40 & \phmm{7.92} & O AGB \\
19 & 76.2801 & $-$65.3911 & 13.72 & 12.09 & 10.72 & \phmm{7.78} & C IR \\
20 & 77.5189 & $-$66.4232 & \nodata & \nodata & \nodata & \phmm{4.12} && K5III & *$|$HD 34127 \\
21 & 76.3348 & $-$66.9185 & 15.58 & 14.60 & 13.61 & \phmm{6.09} & PN && IR$|$IRAS 05053$-$6659 \\
22 & 76.1964 & $-$66.6751 & 14.18 & 13.75 & 12.96 & \phmm{6.06} & PN && IR$|$IRAS 05047$-$6644 \\
23 & 76.1491 & $-$65.7629 & 6.99 & 6.40 & 6.24 & \phmm{6.51} & A-K III & K2III & *$|$HD 33322 \\
24 & 76.4292 & $-$66.8491 & 7.37 & 6.62 & 6.49 & \phmm{6.51} & A-K III & K1 & *$|$HD 33530 \\
25 & 75.7765 & $-$66.9101 & 7.18 & 6.64 & 6.48 & \phmm{6.77} & A-K III & K0III & *iC$|$HD 33148 \\
26 & 75.9988 & $-$65.4274 & 7.76 & 6.99 & 6.85 & \phmm{7.02} & MIII & M0 & *$|$HD 270978 \\
27 & 76.5882 & $-$66.7211 & 16.09 & 13.61 & 11.60 & \phmm{7.02} & OH/IR \\
28 & 77.0320 & $-$65.8135 & 7.80 & 6.94 & 6.63 & \phmm{6.78} & MIII & M & *$|$PV 2088 \\
29 & 77.9111 & $-$66.8528 & 14.32 & 12.38 & 10.74 & \phmm{6.92} & OH/IR \\
30 & 77.9575 & $-$66.4081 & 15.76 & 13.70 & 12.01 & \phmm{7.26} & OH/IR \\
31 & 76.5891 & $-$65.9668 & 7.07 & 6.64 & 6.49 & \phmm{6.78} & A-K III & G8III/IV & **$|$CCDM J05063-6558AB \\
32 & 77.5074 & $-$65.8657 & 15.43 & 13.24 & 11.46 & \phmm{7.11} & OH/IR \\
33 & 76.9190 & $-$66.8159 & 14.49 & 12.50 & 10.93 & \phmm{7.22} \\
34 & 77.1513 & $-$65.5034 & 13.23 & 11.61 & 10.43 & \phmm{7.37} & C IR \\
35 & 76.4690 & $-$66.0135 & 12.30 & 10.88 & 9.70 & \phmm{7.44} & C IR \\
36 & 77.0451 & $-$66.2967 & 8.08 & 7.14 & 6.81 & \phmm{7.03} & MIII & M & *$|$RM 1-206 \\
37 & 77.5041 & $-$65.5461 & 7.70 & 7.20 & 7.03 & \phmm{7.29} & A-K III & K0 & *iC$|$HD 271049 \\
38 & 75.9547 & $-$66.3699 & 14.40 & 12.53 & 10.98 & \phmm{7.20} & OH/IR \\
39 & 76.7218 & $-$65.6687 & 8.26 & 7.68 & 7.52 & \phmm{7.08} & MV & K7 & *iC$|$HD 271012 \\
40 & 75.8935 & $-$65.8078 & 17.06 & 14.51 & 12.35 & \phmm{6.80} & OH/IR \\
41 & 76.5038 & $-$66.2110 & 15.50 & 15.08 & 14.66 & \phmm{7.72} & HII \\
43 & 76.0588 & $-$67.2707 & 8.00 & 7.17 & 6.76 & \phmm{5.33} & O AGB & M4Ia & V*$|$SV* HV 888 \\
44 & 77.7934 & $-$67.8696 & 16.38 & 14.05 & 11.67 & \phmm{5.46} & OH/IR && *$|$IRAS 05112$-$6755 \\
45 & 77.6718 & $-$68.6019 & 16.20 & 14.52 & 11.76 & \phmm{5.78} & OH/IR && IR$|$IRAS 05108$-$6839 \\
46 & 75.9772 & $-$67.3135 & 16.03 & 14.08 & 12.34 & \phmm{6.16} & OH/IR && EmO$|$LHA 120-N 17A \\
47 & 77.8077 & $-$67.6045 & 17.56 & 14.76 & 12.50 & \phmm{6.35} & OH/IR \\
\enddata
\tablenotetext{a}{In decimal degrees.}
\end{deluxetable}


\begin{deluxetable}{cccrrrcccl}
\rotate
\tabletypesize{\small}
\def\phmm{\phm{$-$}}
\tablenum{3}
\tablecolumns{10}
\tablewidth{9.0in}
\tablecaption{MSX Sources in Common with the 2MASS Unreleased Regions\label{Table 3}}
\tablehead{\colhead{MSX \#} & \colhead{RA (J2000)\tablenotemark{a}} & 
\colhead{Dec (J2000)\tablenotemark{a}}
& \colhead{$J$} & \colhead{$H$} & \colhead{$K_{\rm s}$} & \colhead
{A (8.3 $\mu$m)}
& \colhead{IR Type} & \colhead{SIMBAD Type} & \colhead{SIMBAD Name}}
\startdata
271 & 78.9226 & $-$67.9813 & 15.03 & 14.75 & 14.46 & 7.22 & HII     && IR$|$IRAS 05158$-$6802 \\
274 & 79.0249 & $-$66.5650 &  8.12 &  7.62 &  7.48 & 7.67 & A-K III & G0 & *$|$HD 269233 \\
298 & 80.0015 & $-$66.0577 &  6.71 &  6.51 &  6.46 & 6.78 & MS(V) & F2/F3IV/ & *$|$HD 35474 \\
299 & 79.1115 & $-$66.1327 &  8.14 &  7.56 &  7.43 & 7.44 & A-K III & M0 & *$|$HD 271129 \\ 
304 & 79.8203 & $-$67.8636 &  6.41 &  5.56 &  5.33 & 5.34 & MIII & K5III & V*$|$V* AS Dor \\
305 & 79.6362 & $-$67.5422 &  6.56 &  5.81 &  5.55 & 5.69 & MIII & K5 & *$|$HD 35323 \\ 
306 & 79.9574 & $-$66.4431 &  6.46 &  5.86 &  5.68 & 5.82 & A-K III & K1III & *$|$HD 35461 \\
307 & 79.7345 & $-$67.7513 & 17.59 & 15.61 & 12.82 & 6.27 & OH/IR && *$|$IRAS 05190$-$6748 \\
308 & 79.3778 & $-$66.7270 & 14.26 & 13.76 & 13.13 & 6.58 & HII && *$|$TRM 81 \\
309 & 80.0690 & $-$66.8817 & 16.14 & 15.24 & 13.85 & 6.94 &&& Cl*$|$KMHK 786 \\
310 & 80.0807 & $-$66.5966 & 14.38 & 12.36 & 10.86 & 6.59 & OH/IR \\
311 & 79.4210 & $-$66.7041 & 15.90 & 14.54 & 13.76 & 6.84 &&& EmO$|$LHA 120-S 24 \\
312 & 79.9873 & $-$66.2309 & 10.04 &  9.29 &  9.17 & 7.49 && G5 & *$|$HD 269339 \\
313 & 79.0806 & $-$67.2545 &  8.51 &  7.82 &  7.65 & 7.24 & MV & M & *$|$WOH G 252 \\
314 & 80.1444 & $-$66.7797 & 15.37 & 14.64 & 14.48 & 7.11 & HII && *$|$TRM 80 \\
315 & 78.3561 & $-$67.4807 & 16.97 & 15.85 & 14.83 & 7.23 \\
317 & 80.0489 & $-$67.7073 & 17.24 & 15.89 & 16.85 & 7.78 & HII \\
318 & 79.8013 & $-$69.1518 & 15.16 & 14.70 & 12.87 & 5.26 &&& IR$|$IRAS 05195$-$6911 \\
319 & 79.3981 & $-$68.5931 &  5.95 &  5.45 &  5.36 & 5.51 & A-K III & G8III & *$|$HD 35230 \\
320 & 79.8182 & $-$69.3404 & 15.98 & 14.72 & 14.18 & 5.65 & HII && IR$|$LI-LMC 810 \\
321 & 79.7656 & $-$69.4923 & 13.18 & 11.96 & 10.46 & 5.77 &&& V*$|$AGPRS J051904.02$-$69 \\
322 & 79.0447 & $-$69.2447 & 16.33 & 15.34 & 13.40 & 6.07 \\
323 & 79.1325 & $-$68.3692 & 12.19 & 11.68 & 10.53 & 6.05 & PN & A0:Iab: & Em*$|$ARDB 184 \\
324 & 79.8868 & $-$67.8790 &  7.06 &  6.28 &  6.00 & 6.09 & MIII & K5III & *$|$HD 269352 \\
325 & 79.1570 & $-$69.4540 & 15.25 & 12.96 & 11.09 & 6.44 & OH/IR \\
326 & 80.2335 & $-$67.9843 &  6.61 &  6.03 &  5.88 & 6.14 & A-K III & K0III/IV & *iC$|$HD 35665 \\
327 & 79.7781 & $-$68.3601 & 14.56 & 14.19 & 13.56 & 6.62 & HII && EmO$|$LHA 120-N 118 \\
328 & 80.2583 & $-$68.3543 &  7.16 &  6.53 &  6.33 & 6.46 & A-K III & K2III & V*$|$NSV 1957 \\
330 & 79.9719 & $-$68.0677 &  8.38 &  7.59 &  7.20 & 6.70 & MIII & M2Ia & V*$|$SV* HV 2450 \\
331 & 79.4936 & $-$69.2650 &  7.61 &  6.75 &  6.45 & 6.76 & MIII & M & *iA$|$[L72] LH 41-4 \\
332 & 79.2526 & $-$69.3252 & 14.94 & 12.52 & 10.73 & 6.58 & OH/IR && PN$|$Jacoby LMC 17 \\
334 & 79.5092 & $-$69.5374 & 16.60 & 15.40 & 15.05 & 6.80 & HII \\
335 & 79.4070 & $-$69.2550 & 14.40 & 13.46 & 13.41 & 6.84 & HII \\
336 & 79.4850 & $-$69.2455 & 13.96 & 12.94 & 12.62 & 6.78 & HII \\
337 & 79.9945 & $-$68.8955 & 15.73 & 13.68 & 11.86 & 7.02 & OH/IR \\
338 & 79.7347 & $-$67.9372 &  8.63 &  7.86 &  7.47 & 6.85 & RSG \\
339 & 79.0428 & $-$68.8329 & 16.77 & 14.77 & 12.65 & 7.13 & OH/IR \\
341 & 80.2518 & $-$69.3487 & 17.52 & 15.29 & 12.90 & 6.47 & OH/IR \\
342 & 79.7475 & $-$68.7384 &  7.80 &  7.32 &  7.20 & 7.17 & A-K III & G8III & *$|$HD 35447 \\
343 & 79.5135 & $-$68.4658 &  6.95 &  6.54 &  6.42 & 6.85 & A-K III & G2IV: & *iC$|$HD 35294 \\
344 & 79.3441 & $-$69.3376 &  9.80 &  9.32 &  8.81 & 5.97 && ?... & V*$|$HD 35231 \\
345 & 79.6747 & $-$67.9613 & 16.84 & 15.03 & 12.91 & 7.13 & OH/IR \\
346 & 79.6173 & $-$68.0678 & 17.92 & 15.63 & 14.82 & 6.93 \\
347 & 80.0490 & $-$68.6315 & 16.48 & 15.06 & 15.28 & 7.35 & HII \\
348 & 79.7875 & $-$69.4555 & 14.03 & 12.22 & 10.75 & 7.53 & C IR \\
\enddata
\tablenotetext{a}{In decimal degrees.}
\end{deluxetable}

\begin{deluxetable}{cccccccclccc}
\ptlandscape
\rotate
\tabletypesize{\small}
\def\phmm{\phm{$-$}}
\tablenum{4}
\tablecolumns{12}
\tablewidth{9.0in}
\tablecaption{MSX Sources Without 2MASS Point Source Matches\label{Table 4}}
\tablehead{\colhead{MSX \#} & \colhead{RA\tablenotemark{a}} & 
\colhead{Dec\tablenotemark{a}} & \colhead{A} &
\colhead{B} &  \colhead{C} &  \colhead{D} &  \colhead{SIMBAD} &
\colhead{SIMBAD} & \multicolumn{3}{c}{Notes\tablenotemark{b}} \\
\colhead{} & \colhead{(J2000)} & \colhead{(J2000)} & 
\colhead{(8.3 $\mu$m)} & \colhead{(12.1 $\mu$m)} & 
\colhead{(14.7 $\mu$m)} & \colhead{(21.3 $\mu$m)} & 
\colhead{Type} & \colhead{Name} & \colhead{N1} & \colhead{N2} & 
\colhead{N3}}
\startdata
  17 & 77.5329 & $-$64.3681 &   7.15 & \nodata & \nodata & \nodata  	&&&	  3     \\ 
  42 & 75.3938 & $-$68.0987 &   2.79 &   2.12 &   1.94 &   1.77  	& M7 & Mi*$|$V* RX Dor &	  5      \\ 
  58 & 75.4509 & $-$67.7917 &   7.00 & \nodata & \nodata & \nodata  	&&&	  4&& 4   \\ 
  63 & 76.2797 & $-$68.0524 &   7.22 & \nodata & \nodata & \nodata  	&& *$|$[O96] D066 - 170 &	  2 & 3 \\ 
  84 & 76.2687 & $-$68.9635 &   5.90 & \nodata & \nodata & \nodata  	&&&	  4      \\ 
  86 & 76.2171 & $-$70.1256 &   6.30 & \nodata & \nodata & \nodata  	&& IRAS 05052$-$7011&  5 & 3 \\ 
 133 & 76.1563 & $-$70.9103 &   5.26 &   3.40 &   3.58 &   0.59  	& ?... & EmO$|$LHA 120-N 191A &	  4& 3& 5\\ 
 145 & 76.1065 & $-$70.7265 &   6.62 & \nodata & \nodata & \nodata  	&&&	  5 &  2\\ 
 185 & 74.6938 & $-$73.2010 &   7.37 & \nodata & \nodata & \nodata  	&&&	  2 &  5\\ 
 187 & 77.9137 & $-$73.7902 &   9.80 & \nodata & \nodata & \nodata  	&&&	  2 &  5\\ 
 272 & 78.4541 & $-$67.1250 &   7.32 & \nodata & \nodata & \nodata  	&&&	  2      \\ 
 282 & 78.8922 & $-$65.5438 &   2.42 &   2.31 &   2.21 & \nodata  	& F0 & *$|$HD 271114 &	  4&& 3   \\ 
 284 & 78.5873 & $-$65.3136 &   7.53 & \nodata & \nodata & \nodata  	&&&	  2 &  5\\ 
 288 & 78.4921 & $-$66.4102 &   7.63 & \nodata & \nodata & \nodata  	&&&	  3 &  4\\ 
 302 & 78.6732 & $-$65.6551 &   7.37 & \nodata & \nodata & \nodata  	&&&	  3 &  4\\ 
 329 & 79.6844 & $-$69.2407 &   6.63 & \nodata & \nodata & \nodata  	&& *iA$|$[L72] LH 41-1095&  2 &  3\\ 
 377 & 77.8470 & $-$70.3696 &   7.25 & \nodata & \nodata & \nodata  	&&&	  2 &  3\\ 
 404 & 77.5649 & $-$72.6087 &   7.03 & \nodata & \nodata & \nodata  	&&&	  2 &  2\\ 
 407 & 79.0652 & $-$71.5650 &   7.40 & \nodata & \nodata & \nodata  	&&&	  4 &  3\\ 
 410 & 78.1202 & $-$72.7943 &   7.59 & \nodata & \nodata & \nodata  	&&&	  3& 4& 4\\ 
 414 & 79.5650 & $-$72.9557 &   7.32 & \nodata & \nodata & \nodata  	&&&	  2 &  5\\ 
 415 & 79.9805 & $-$73.1202 &   7.30 & \nodata & \nodata & \nodata  	&&&	  3 &  4\\ 
 437 & 80.8134 & $-$69.9353 &   6.24 & \nodata & \nodata & \nodata  	&&&	  4      \\ 
 463 & 79.7829 & $-$69.6295 &   5.59 & \nodata & \nodata & \nodata  	&&&	  2      \\ 
 509 & 80.7018 & $-$68.0194 &   6.87 & \nodata & \nodata & \nodata  	& & EmO$|$LHA 120-N 44H  &	  4&& 2   \\ 
 527 & 80.5818 & $-$65.7217 &   6.39 &   3.71 & \nodata & \nodata  	& & *$|$TRM 107 &	  4&& 5   \\ 
 573 & 81.9170 & $-$67.4465 &   7.06 & \nodata & \nodata & \nodata  	&&&	  4 &  3\\ 
 576 & 80.7934 & $-$66.3788 &   7.38 & \nodata & \nodata & \nodata  	&&&	  3& 3& 2\\ 
 584 & 80.5128 & $-$67.9633 &   4.80 & \nodata & \nodata & \nodata  	& & Rad$|$MRC 0522$-$680 &	  2& 3& 2\\ 
 691 & 82.7145 & $-$72.7612 &   6.61 & \nodata & \nodata & \nodata  	&&&	  4      \\ 
 715 & 83.7326 & $-$73.6288 &   7.45 & \nodata & \nodata & \nodata  	&&&	  3 &  4\\ 
 723 & 82.4082 & $-$72.8314 &   6.99 & \nodata & \nodata & \nodata  	&&&	  2&& 2   \\ 
 726 & 82.5995 & $-$72.8924 &   7.40 & \nodata & \nodata & \nodata  	&&&	  2& 4 &4\\ 
 739 & 83.3772 & $-$69.8767 &   6.78 & \nodata & \nodata & \nodata  	& & IR$|$IRAS 05339$-$6954     &	  4 &  2\\ 
 761 & 83.8210 & $-$71.2972 &   7.69 & \nodata & \nodata & \nodata  	&&&	  3 &  4\\ 
 763 & 83.5603 & $-$69.7891 &   3.93 &   1.92 &   0.80 &  $-$0.36  	& & IR$|$IRAS 05346$-$6949 &	  5      \\ 
 800 & 82.8883 & $-$68.4788 &   7.49 & \nodata & \nodata & \nodata  	&&&	  2& 3& 5\\ 
 808 & 83.2154 & $-$67.6843 &   6.08 & \nodata & \nodata &   1.81  	& & *$|$[O96] D231 - 263 &	  5& 3& 2\\ 
 842 & 82.4906 & $-$65.8392 &   7.25 & \nodata & \nodata & \nodata  	& & *$|$TRM 103 &	  2      \\ 
 849 & 82.1467 & $-$66.5047 &   7.32 & \nodata & \nodata & \nodata  	&&&	  2&& 5   \\ 
 855 & 82.9190 & $-$64.5423 &   7.23 & \nodata & \nodata & \nodata  	&&&	  2& 5& 5\\ 
 859 & 83.8405 & $-$63.9374 &   7.37 & \nodata & \nodata & \nodata  	&&&	  2 &  5\\ 
 865 & 84.1946 & $-$65.7923 &   7.49 & \nodata & \nodata & \nodata  	&&&	  3      \\ 
 868 & 83.8485 & $-$67.5824 &   4.82 &   3.24 &   1.98 &   0.01  	& ?e... & *$|$HD 37731 &	  4& 3& 2\\ 
 895 & 84.4955 & $-$69.1082 &   6.23 & \nodata & \nodata &   1.24  	&&&	  3      \\ 
 898 & 84.8831 & $-$69.0980 &   6.59 & \nodata & \nodata & \nodata  	&&&	  3& 3& 2\\ 
 902 & 84.6697 & $-$69.0020 &   6.92 & \nodata & \nodata & \nodata  	&&&	  2& 3& 2\\ 
 920 & 84.7287 & $-$69.1769 &   7.09 & \nodata & \nodata & \nodata  	&&&	  3 &  3\\ 
 933 & 84.9117 & $-$69.7704 &   4.41 &   2.63 &   1.73 &  $-$0.43  	& &*$|$[DCL92] 220 &	  4& 3& 2\\ 
 945 & 84.0293 & $-$69.5411 &   6.73 & \nodata & \nodata & \nodata  	&&&	  3 &  3\\ 
 953 & 84.7232 & $-$69.5713 &   6.85 & \nodata & \nodata & \nodata  	&&&	  2 &  3\\ 
 964 & 84.3330 & $-$69.7533 &   7.23 & \nodata & \nodata & \nodata  	&&&	  2 &  3\\ 
 980 & 85.6505 & $-$71.5931 &   6.78 & \nodata & \nodata & \nodata  	&&&	  3      \\ 
 989 & 83.5356 & $-$71.9283 &   7.26 & \nodata & \nodata & \nodata  	&&&	  2      \\ 
 996 & 66.6111 & $-$71.9164 &   7.55 & \nodata & \nodata & \nodata  	&&&	  2 &  4\\ 
1002 & 66.2966 & $-$71.7966 &   7.89 & \nodata & \nodata & \nodata  	&&&	  3 &  4\\ 
1041 & 70.2596 & $-$66.3808 &   7.57 & \nodata & \nodata & \nodata  	&&&	  2 &  5\\ 
1059 & 66.7514 & $-$73.1858 &   7.40 & \nodata & \nodata & \nodata  	&&&	  2 &  4\\ 
1060 & 67.6442 & $-$72.7854 &   7.81 & \nodata & \nodata & \nodata  	&&&	  3 &  2\\ 
1062 & 68.4079 & $-$73.2724 &   7.07 & \nodata & \nodata & \nodata  	&&&	  4      \\ 
1087 & 72.0591 & $-$66.5192 &   7.53 & \nodata & \nodata & \nodata  	&&&	  2 &  4\\ 
1094 & 72.7494 & $-$64.0787 &   7.34 & \nodata & \nodata & \nodata  	&&&	  2 &  4\\ 
1111 & 73.4342 & $-$66.1963 &   7.15 & \nodata & \nodata & \nodata  	& &IR$|$IRAS 04535$-$6616 &	  4      \\ 
1114 & 72.9071 & $-$66.3558 &   7.63 & \nodata & \nodata & \nodata  	&&&	  3 &  5\\ 
1153 & 69.6848 & $-$72.8560 &   7.53 & \nodata & \nodata & \nodata  	&&&	  2 &  4\\ 
1154 & 69.5942 & $-$72.8798 &   7.49 & \nodata & \nodata & \nodata  	&&&	  3 &  4\\ 
1157 & 69.8409 & $-$73.3907 &   7.63 & \nodata & \nodata & \nodata  	&&&	  3 &  5\\ 
1158 & 71.9523 & $-$73.4623 &   7.44 & \nodata & \nodata & \nodata  	&&&	  2 &  5\\ 
1168 & 71.6976 & $-$71.8944 &   7.63 & \nodata & \nodata & \nodata  	&&&	  3 &  4\\ 
1224 & 73.5865 & $-$67.2854 &   4.29 & \nodata & \nodata & \nodata  	&M1III & *$|$HD 31907 &	  4&& 3   \\ 
1231 & 74.4139 & $-$66.4537 &   6.37 & \nodata & \nodata & \nodata  	&&&	  4& 3& 2\\ 
1232 & 74.1159 & $-$66.5335 &   6.45 & \nodata & \nodata &   1.35  	&&&	  4& 3& 2\\ 
1245 & 73.4059 & $-$66.9893 &   7.59 & \nodata & \nodata & \nodata  	& & As*$|$BSDL 164 &	  3 &  3\\ 
1255 & 73.8037 & $-$65.5139 &   7.27 & \nodata & \nodata & \nodata  	&&&	  2 &  4\\ 
1264 & 74.1492 & $-$64.7583 &   7.67 & \nodata & \nodata & \nodata  	&&&	  3 &  4\\ 
1279 & 74.1375 & $-$66.4000 &   6.39 & \nodata & \nodata & \nodata  	& &*$|$PGMW 3010 &	  2& 3& 2\\ 
1288 & 74.5009 & $-$66.3388 &   7.35 & \nodata & \nodata & \nodata  	&&&	  2 &  4\\ 
1300 & 74.4323 & $-$68.9998 &   6.08 & \nodata & \nodata & \nodata  	&K3III:&*$|$HD 32439&	  4& 2& 4\\ 
1301 & 74.3282 & $-$68.4229 &   6.24 & \nodata & \nodata & \nodata  	& &As*$|$LH 12 &	  4& 2& 2\\ 
1303 & 74.6752 & $-$68.1208 &   6.33 & \nodata & \nodata & \nodata  	& &IR$|$IRAS 04588$-$6811 &	  4&& 2   \\ 
1313 & 74.6771 & $-$69.2810 &   6.94 & \nodata & \nodata & \nodata  	&&&	  4      \\ 
1347 & 73.2246 & $-$72.3514 &   6.67 & \nodata & \nodata & \nodata  	&&&	  4&& 2   \\ 
1364 & 74.8847 & $-$66.5081 &   7.32 & \nodata & \nodata & \nodata  	&&&	  3      \\ 
1368 & 75.8687 & $-$66.1220 &   7.37 & \nodata & \nodata & \nodata  	&&&	  2 &  5\\ 
1372 & 87.2271 & $-$72.4007 &   7.00 & \nodata & \nodata & \nodata  	&&&	  4&& 5   \\ 
1374 & 87.0527 & $-$72.4243 &   7.83 & \nodata & \nodata & \nodata  	&&&	  3 &  5\\ 
1376 & 86.8816 & $-$73.2954 &   7.34 & \nodata & \nodata & \nodata  	&&&	  2 &  5\\ 
1391 & 85.5634 & $-$68.2017 &   6.50 & \nodata & \nodata & \nodata  	& &IR$|$IRAS 05424$-$6813 &	  4      \\ 
1394 & 85.9433 & $-$67.4572 &   7.06 & \nodata & \nodata & \nodata  	& &*$|$TRM 35 &	  4& 3& 2\\ 
1421 & 85.2201 & $-$64.2655 &   7.13 & \nodata & \nodata & \nodata  	&&&	  2      \\ 
1423 & 86.0165 & $-$65.8995 &   6.44 & \nodata & \nodata & \nodata  	& &IR$|$IRAS 05439$-$6555 &	  4      \\ 
1425 & 84.8841 & $-$66.0179 &   7.15 & \nodata & \nodata & \nodata  	&&&	  2 &  5\\ 
1426 & 85.3214 & $-$65.9173 &   7.44 & \nodata & \nodata & \nodata  	&&&	  3 &  5\\ 
1427 & 85.5904 & $-$64.8141 &   7.76 & \nodata & \nodata & \nodata  	&&&	  3 &  5\\ 
1450 & 85.7702 & $-$68.0107 &   7.25 & \nodata & \nodata & \nodata  	&&&	  2      \\ 
1469 & 87.6096 & $-$69.9341 &   6.68 & \nodata & \nodata & \nodata  	&&&	  2      \\ 
1496 & 87.8891 & $-$71.3262 &   7.20 & \nodata & \nodata & \nodata  	& &IR$|$IRAS 05522$-$7120 &	  4      \\ 
1510 & 89.6105 & $-$73.3487 &   7.44 & \nodata & \nodata & \nodata  	&&&	  3 &  4\\ 
1518 & 90.5301 & $-$72.4556 &   6.04 & \nodata & \nodata & \nodata  	&&&	  4      \\ 
1558 & 87.7577 & $-$66.3156 &   7.27 & \nodata & \nodata & \nodata  	&&&	  2 &  5\\ 
1559 & 87.2860 & $-$65.6227 &   7.63 & \nodata & \nodata & \nodata  	&&&	  3 &  3\\ 
1562 & 86.7848 & $-$64.6343 &   7.57 & \nodata & \nodata & \nodata  	&&&	  3& 5& 2\\ 
1565 & 88.6607 & $-$64.9388 &   3.29 &   2.91 &   2.75 &   1.81  	&M & *$|$WOH G 622 &	  4&& 3   \\ 
1574 & 88.2760 & $-$65.8905 &   7.34 & \nodata & \nodata & \nodata  	&&&	  2 &  4\\ 
1585 & 87.7560 & $-$66.6313 &   7.23 & \nodata & \nodata & \nodata  	&&&	  3&& 4   \\ 
1588 & 87.1242 & $-$66.1807 &   7.59 & \nodata & \nodata & \nodata  	&&&	  3 &  5\\ 
1592 & 89.1615 & $-$67.8929 &   5.81 & \nodata & \nodata & \nodata  	& &C*$|$IRAS 05568$-$6753 &	  4      \\ 
1601 & 89.6088 & $-$69.7401 &   7.06 & \nodata & \nodata & \nodata  	& &IR$|$IRAS 05588$-$6944 &	  4&& 4   \\ 
1603 & 90.1559 & $-$69.7474 &   7.07 & \nodata & \nodata & \nodata  	&&&	  2 &  4\\ 
1606 & 88.6677 & $-$70.2485 &   7.45 & \nodata & \nodata & \nodata  	&&&	  3 &  5\\ 
1612 & 91.4862 & $-$72.6481 &   4.10 &   3.31 & \nodata & \nodata  	&M4 &*$|$HD 271928 &	  4&& 3   \\ 
1615 & 90.2007 & $-$73.2832 &   7.16 & \nodata & \nodata & \nodata  	&&&	  2& 5& 2\\ 
1616 & 91.4891 & $-$73.3671 &   7.40 & \nodata & \nodata & \nodata  	&&&	  3 &  4\\ 
1617 & 92.1600 & $-$73.3898 &   7.44 & \nodata & \nodata & \nodata  	&&&	  3 &  4\\ 
1623 & 94.0324 & $-$73.1784 &   7.20 & \nodata & \nodata & \nodata  	&&&	  2 &  4\\ 
1630 & 91.6009 & $-$71.2280 &   7.32 & \nodata & \nodata & \nodata  	&&&	  3 &  5\\ 
1634 & 93.1559 & $-$71.9089 &   7.44 & \nodata & \nodata & \nodata  	&&&	  3 &  4\\ 
1647 & 89.0121 & $-$67.9521 &   7.23 & \nodata & \nodata & \nodata  	&&&	  2 &  5\\ 
1648 & 91.5333 & $-$69.1539 &   7.29 & \nodata & \nodata & \nodata  	&&&	  2 &  5\\ 
1649 & 90.5638 & $-$69.5798 &   7.38 & \nodata & \nodata & \nodata  	&&&	  3 &  4\\ 
1654 & 89.6045 & $-$66.9275 &   7.16 & \nodata & \nodata & \nodata  	&&&	  2 &  5\\ 
1657 & 88.4999 & $-$67.0307 &   7.26 & \nodata & \nodata & \nodata  	&&&	  2 &  4\\ 
1661 & 90.6877 & $-$67.5763 &   7.35 & \nodata & \nodata & \nodata  	&&&	  3 &  4\\ 
1664 & 88.7482 & $-$65.9538 &   7.35 & \nodata & \nodata & \nodata  	&&&	  2 &  4\\ 
1683 & 90.1857 & $-$64.9369 &   7.23 & \nodata & \nodata & \nodata  	&&&	  2& 3& 2\\ 
1693 & 90.4521 & $-$66.9300 &   7.26 & \nodata & \nodata & \nodata  	&&&	  2 &  5\\ 
1712 & 92.1238 & $-$69.1775 &   7.51 & \nodata & \nodata & \nodata  	&&&	  3 &  5\\ 
1713 & 92.9640 & $-$69.8555 &   7.67 & \nodata & \nodata & \nodata  	&&&	  3 &  4\\ 
1718 & 93.3395 & $-$71.2765 &   7.42 & \nodata & \nodata & \nodata  	&&&	  3 &  4\\ 
1720 & 94.9976 & $-$72.7518 &   4.09 & \nodata & \nodata & \nodata  	&M2 &*$|$HD 272138 &	  4&& 3   \\ 
1733 & 95.2990 & $-$72.2156 &   7.51 & \nodata & \nodata & \nodata  	&&&	  3 &  5\\ 
1740 & 93.3803 & $-$70.7388 &   7.25 & \nodata & \nodata & \nodata  	&&&	  2 &  4\\ 
1762 & 71.8456 & $-$71.2150 &   7.29 & \nodata & \nodata & \nodata  	&&&	  2 &  5\\ 
1764 & 74.0195 & $-$70.4696 &   7.30 & \nodata & \nodata & \nodata  	&&&	  2 &  5\\ 
1769 & 76.7236 & $-$65.7881 &   7.22 & \nodata & \nodata & \nodata  	&&&	  2 &  4\\ 
1770 & 77.9071 & $-$64.2270 &   7.04 & \nodata & \nodata & \nodata  	&&&	  2 &  3\\ 
1773 & 78.4313 & $-$72.3520 &   7.04 & \nodata & \nodata & \nodata  	&&&	  2 &  4\\ 
1783 & 82.7823 & $-$73.0388 &   7.11 & \nodata & \nodata & \nodata  	&&&	  2 &  4\\ 
1790 & 84.2975 & $-$66.8662 &   7.01 & \nodata & \nodata & \nodata  	&&&	  2      \\ 
1792 & 83.9179 & $-$64.1715 &   6.92 & \nodata & \nodata & \nodata  	&&&	  2      \\ 
1796 & 84.9444 & $-$69.2161 &   7.04 & \nodata & \nodata & \nodata  	&&&	  2 &  3\\ 
1800 & 86.5976 & $-$72.3855 &   7.26 & \nodata & \nodata & \nodata  	&&&	  2 &  3\\ 
\enddata
\tablenotetext{a}{In decimal degrees.}
\tablenotetext{b}{Explanation of Notes. N1 is the number of MSX scan legs in 
which the source was detected, which is a possible measure of the source's 
reliability.
N2 is a code for the source appearance in the MSX A-band coadded image:
blank = point source in MSX image; 2 = multiple/blended source at 
position; 3 = extended emission source at position;
4 = very low SNR source, possibly not real; and, 5 = no apparent emission.  
N3 is a code for the source appearance in the 2MASS images:
blank = source not seen in 2MASS images; 2 = extended 2MASS emission;
3 = very bright (saturated) 2MASS star; 4 = possible positional mismatch; and,
5 = possible faint emission at source position in 2MASS images.}
\end{deluxetable}

\clearpage

\begin{figure}[tbp]
\figurenum{1} \plotone{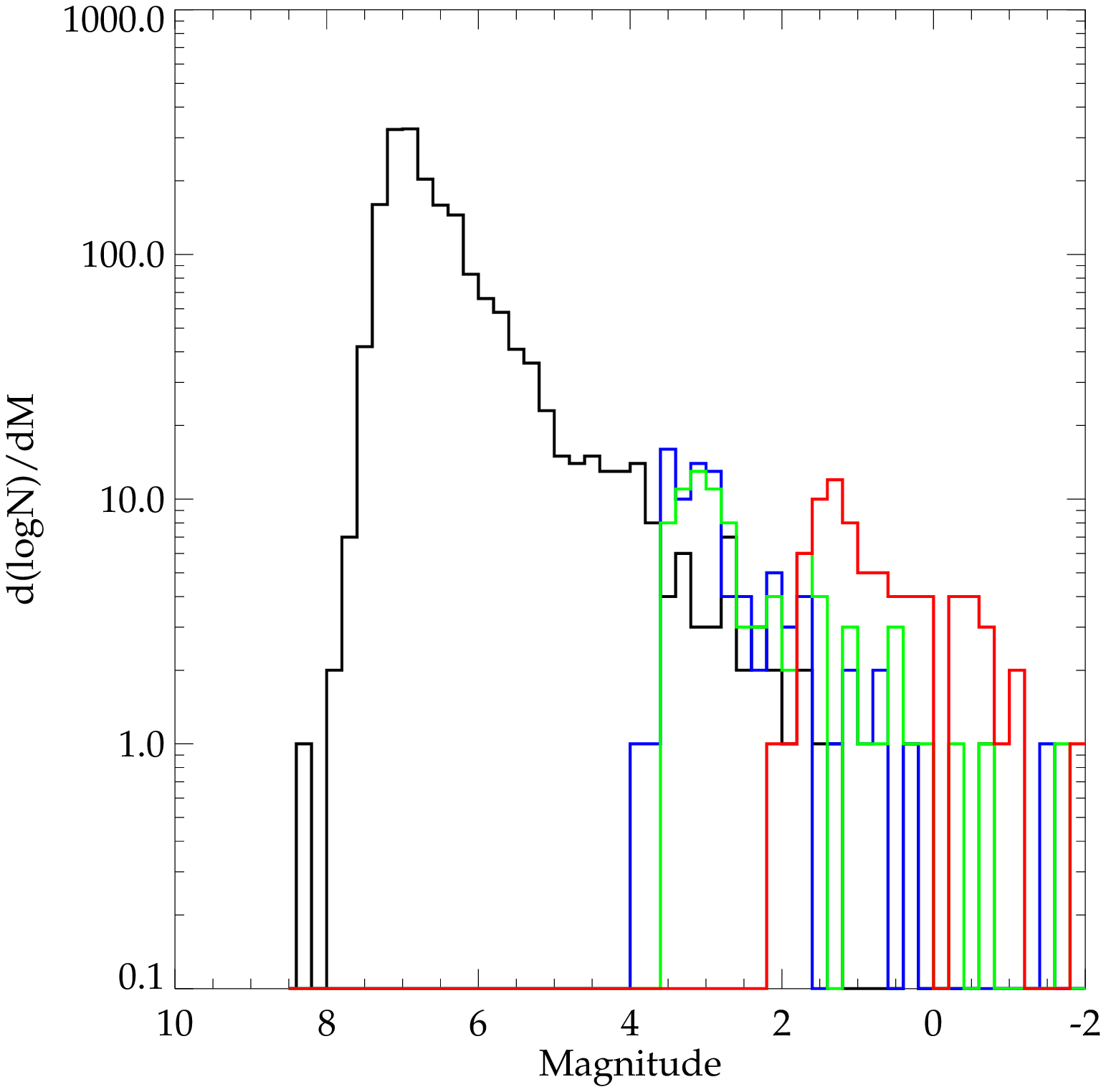}
\caption{The distribution of source brightness, $d{\log }N/dM$, as a
function of MSX magnitude for bands A (8.3 $\protect\mu $m; {\it black 
line}), C (12.1 $\protect\mu $m; {\it blue line}), D (14.6 $\protect%
\mu $m; {\it green line}), and E (21.3 $\protect\mu $m; {\it red 
line}).}
\end{figure}

\clearpage

\begin{figure}[tbp]
\figurenum{2} \plotone{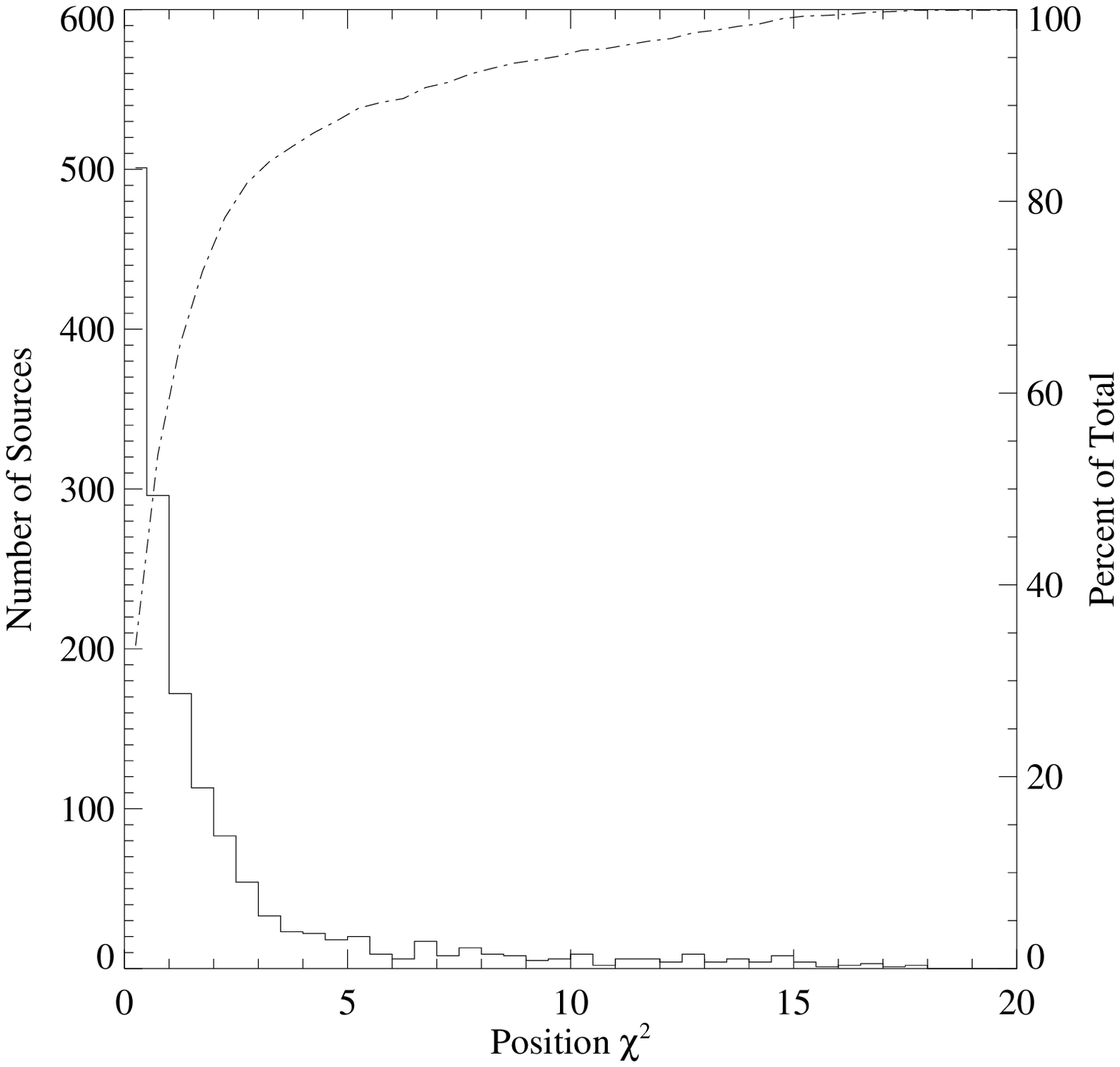}
\caption{The distribution of the position error goodness-of-fit parameter $%
\protect\chi ^{2}$ values for matches between the MSX PSC and the 2IDR PSC (%
{\it solid line} and leftmost ordinate). The {\it dot-dashed line\/} (and
rightmost ordinate) shows the cumulative fraction of source matches as a
function of $\protect\chi ^{2}$.}
\end{figure}

\begin{figure}[tbp]
\figurenum{3} 
\caption{Color-color diagram of $J-K_{s}$ {\it vs.\/} $K_{s}-$A, where the
colors of the points (see color bar) represents $H-K_{s}$. The uncertainties
in the colors are also represented by the error bars around each point.}
\end{figure}

\clearpage

\begin{figure}[tbp]
\figurenum{4} \plotone{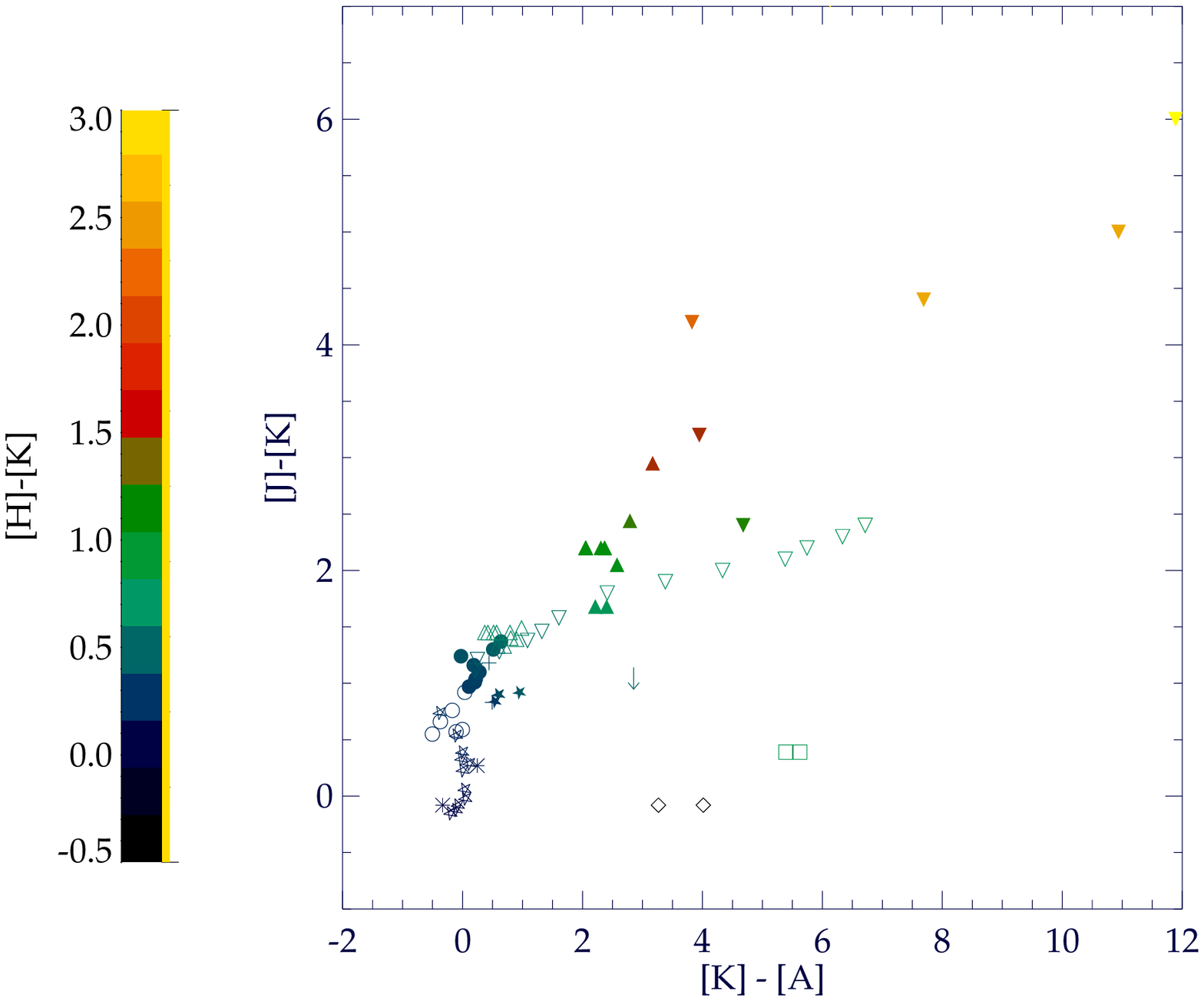}
\caption{Color-color diagram generated from the Wainscoat et al.~(1992)
``SKY'' model of the Galaxy. This figure is the model analog to Figure 3. O
and B dwarfs are represented by {\it asterisks}; A -- K dwarfs, {\it open
five-pointed stars}; M dwarfs, {\it filled five-pointed stars}; A -- G
giants, {\it open circles}; K and M giants, {\it filled circles};
supergiants, {\it plus signs}; C-rich AGB stars (low mass-loss rate), {\it %
open triangles}; C-rich AGB stars (IR carbon stars, high mass-loss rate), 
{\it filled triangles}; O-rich AGB stars (low mass-loss rate), {\it open
upside-down triangles}; OH/IR stars, O-rich hypergiants (high mass-loss
rate), {\it filled upside-down triangles}; PNe, {\it open squares}; T Tauri
objects, {\it down arrow}; and, reflection nebulae, {\it open diamonds}.}
\end{figure}

\begin{figure}[tbp]
\figurenum{5} 
\caption{The $K_{s}-$A {\it vs.\/} $K_{s}$ color-magnitude diagram. The
points have been color-coded to reflect their $J-K_{s}$ colors (see color
bar), so they can be easily referenced to the color-color diagram in Figure
3. The {\it dot-dashed line\/} is the limit represented by the depth of the
MSX survey in band A, approximately magnitude 8. The {\it black solid line\/}
is the 2MASS $K_{s}$ magnitude at SNR=10. The {\it blue solid line\/} is the
projected {\sl SIRTF\/} IRAC instrument sensitivity (see \S 4.6).}
\end{figure}

\clearpage

\begin{figure}[tbp]
\figurenum{6} \plotone{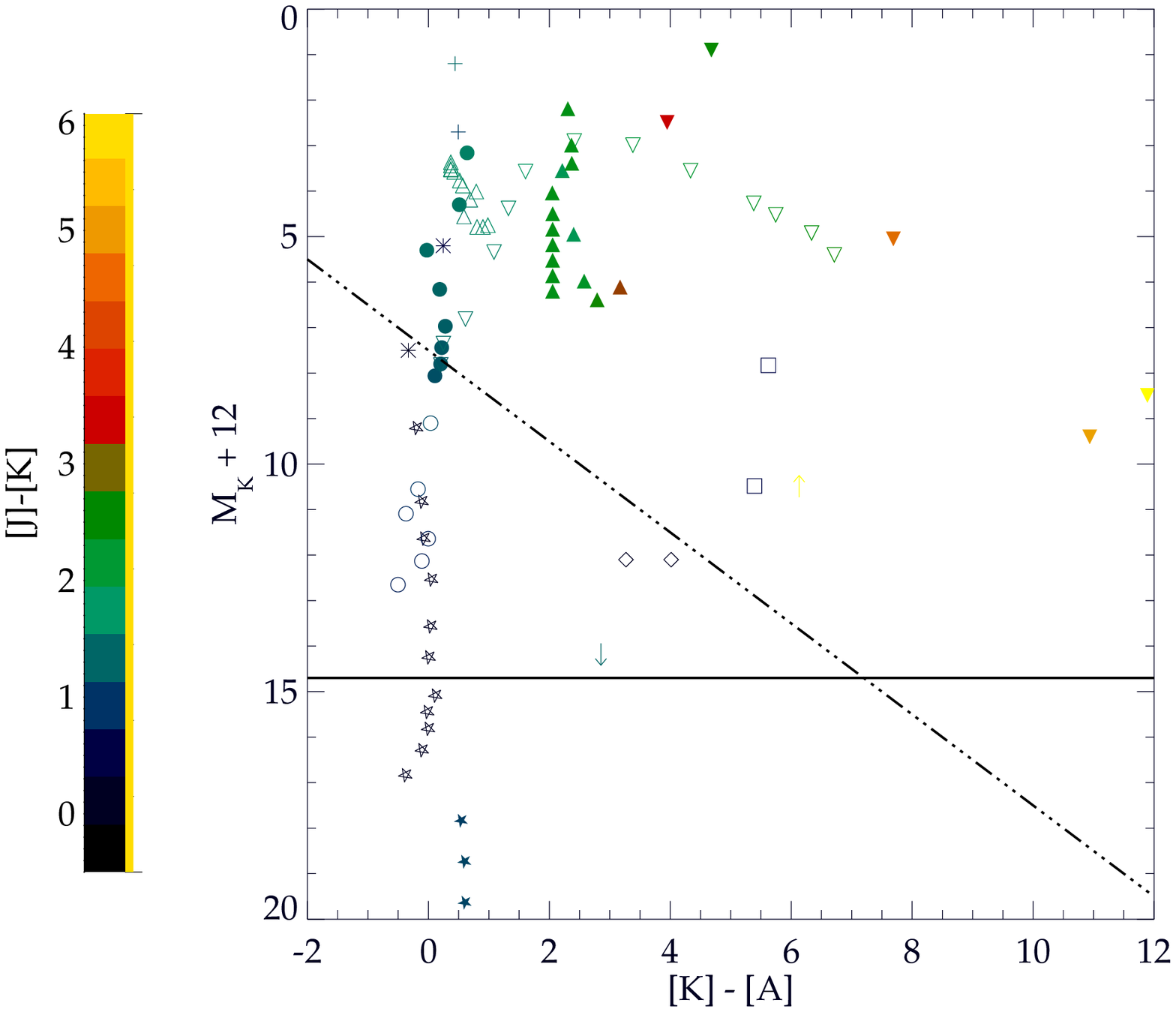}
\caption{The color-magnitude diagram generated from the SKY model from
Wainscot et al.~(1992), computing $K_{s}$ from the model absolute magnitude, 
$M_{K}$, and distance modulus, $\protect\mu =12$, corresponding to
foreground sources. As in Figure 5, the {\it dot-dashed line\/} is the limit
represented by the depth of the MSX survey in band A, and the {\it black
solid line\/} is the 2MASS $K_{s}$ magnitude at SNR=10.}
\end{figure}

\clearpage

\begin{figure}[tbp]
\figurenum{7} \plotone{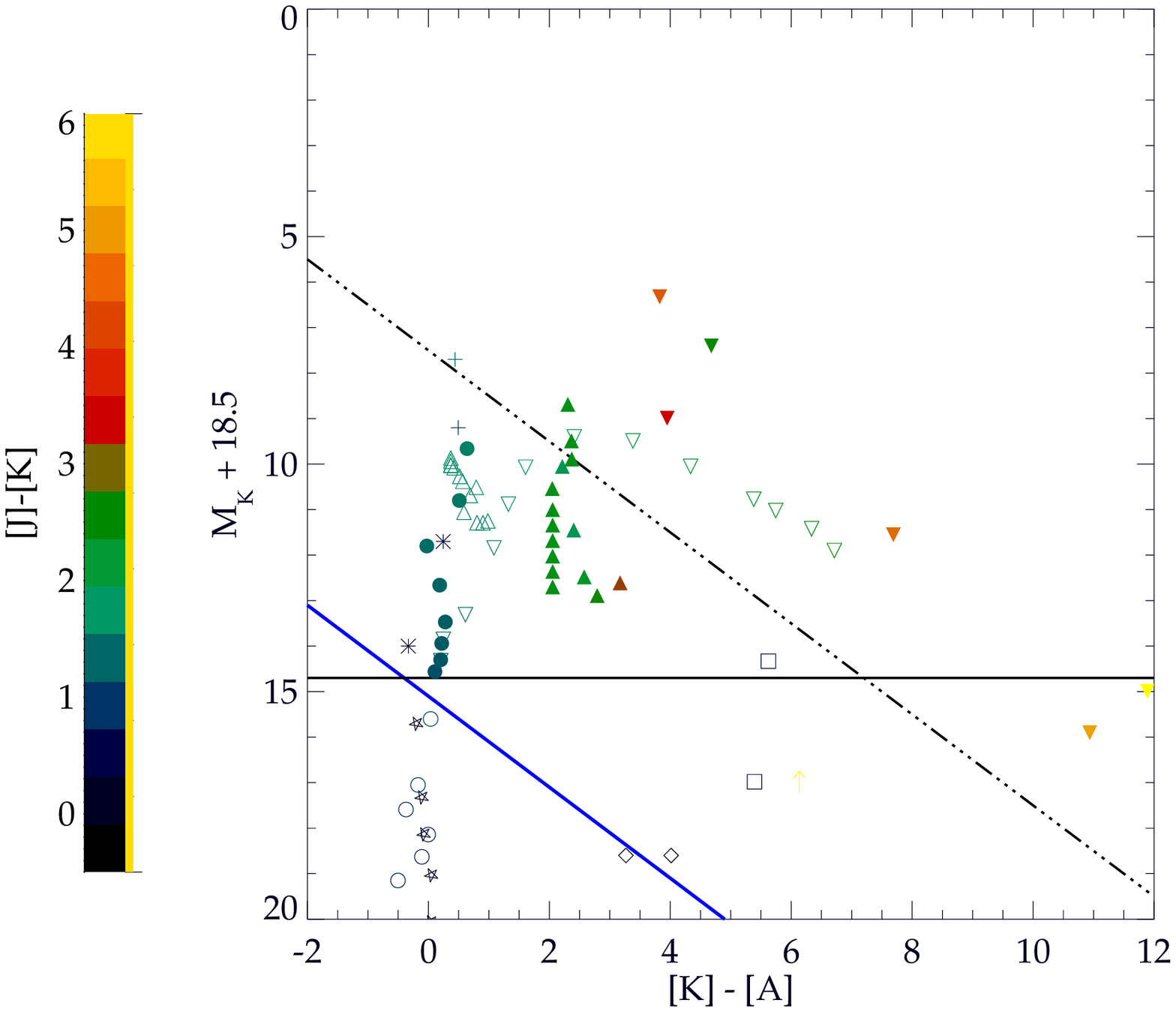}
\caption{The color-magnitude diagram generated from the SKY model of
Wainscoat et al.~(1992), computing $K_{s}$ from the model absolute
magnitude, $M_{K}$, and distance modulus $\protect\mu =18.5$ (see text),
corresponding to LMC sources. As in Figure 5, the {\it dot-dashed line\/} is
the limit represented by the depth of the MSX survey in band A, and the {\it %
blue solid line\/} is the projected {\sl SIRTF\/} IRAC instrument
sensitivity (see \S 4.6).}
\end{figure}

\begin{figure}[tbp]
\figurenum{8} 
\caption{(a) The spatial distribution of sources with luminosity class III
and V, superposed on the MSX A-band mosaic of the LMC. M dwarf stars (class
V) are represented by {\it blue circles}; all other dwarfs are {\it green
circles}. G- and K-type giant stars are {\it red circles}, and M-type giants
are {\it yellow circles}. These sources are all most likely in the Galactic
foreground.  (b) The spatial distribution of the remaining sources, superposed
on the MSX A-band mosaic of the LMC. OH/IR stars are represented by {\it red
diamonds}; IR-bright carbon stars, {\it cyan diamonds}; PNe, {\it magenta
triangles}; HII regions, {\it blue triangles}; RSG stars, {\it green diamonds%
}; O-rich AGB stars, {\it red squares}; and, C-rich AGB stars, {\it cyan
squares}.}
\end{figure}

\clearpage

\begin{figure}[tbp]
\figurenum{9} \plotone{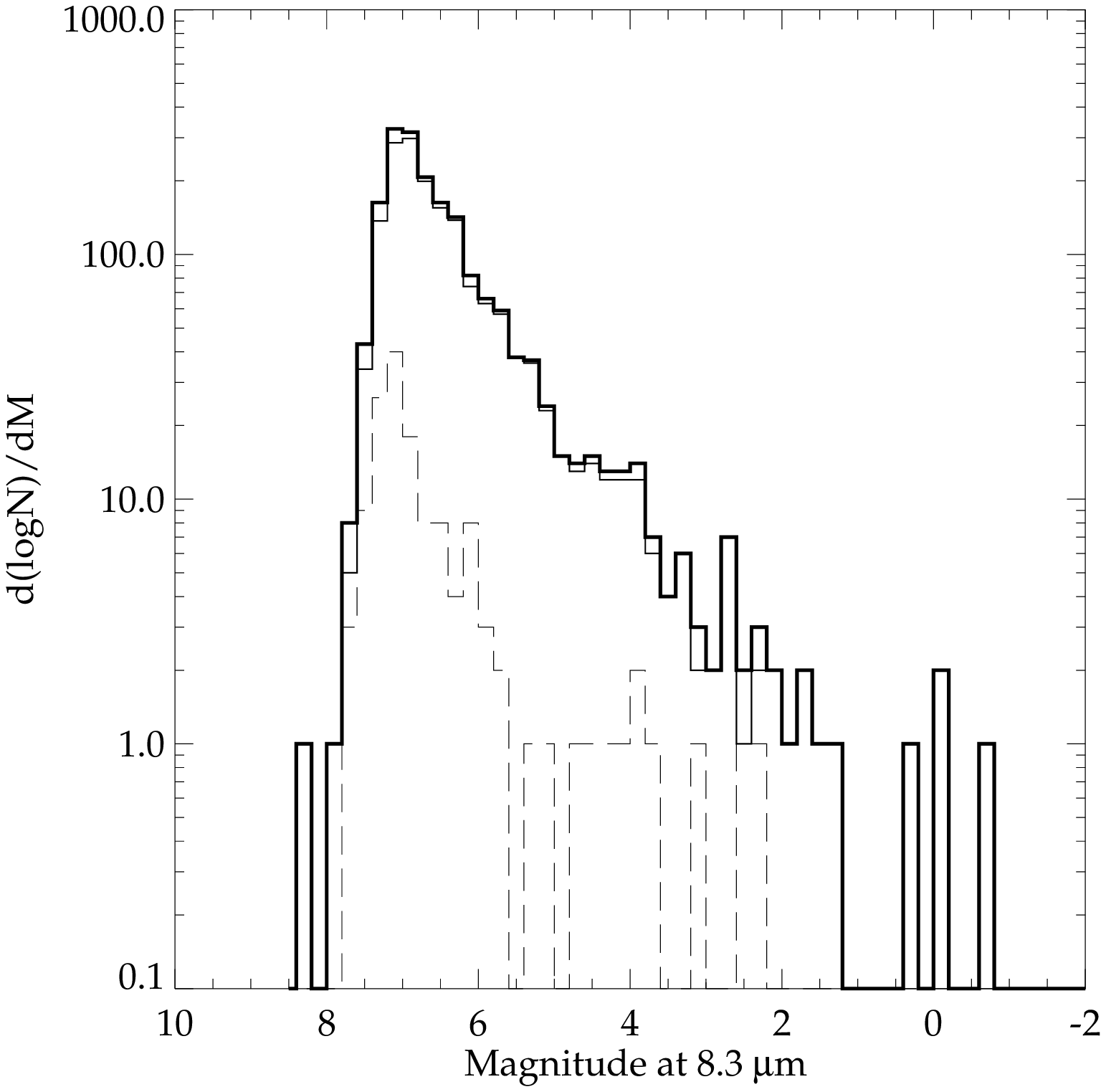}
\caption{The magnitude distribution of MSX band A sources not matched with
2MASS sources ({\it dashed line}). For comparison, the full catalog magnitude
distribution ({\it thick solid line}) and the distribution of matched sources
({\it thin solid line}) are also shown.}
\end{figure}

\end{document}